\documentclass[
 reprint,
 amsmath,amssymb,
 aps,
 showkeys
]{revtex4-1}

\usepackage{graphicx}
\usepackage{dcolumn}
\usepackage{bm}
\usepackage{siunitx}
\usepackage{subfigure}
\usepackage[T1]{fontenc}
\usepackage{float}
\usepackage{xcolor}
\usepackage{soul}
\usepackage{systeme}
\usepackage{wrapfig}
\usepackage{svg}
\usepackage{caption}
\usepackage[hidelinks]{hyperref}
\usepackage{scrlayer-scrpage}
\usepackage{amsmath}
\usepackage{makecell}

\hypersetup{
  colorlinks   = true,
  urlcolor     = blue,
  linkcolor    = blue, 
  citecolor   = blue 
}

\makeatother

\makeatletter
\let\newfloat\newfloat@ltx
\makeatother
\usepackage{algorithm}
\usepackage{algpseudocode}

\usepackage{multirow}
\usepackage{hhline}

\begin{document}

\title{Physically Consistent 2D to 3D Pore Space Synthesis: 1. Dynamic Grain Packing Under Strict Morphology Constraints}

\author{Dmitry Kulygin}
\author{Andrey Ananev}
\author{Aleksey Khlyupin}
\email{khlyupin@phystech.edu}
\affiliation{Laboratory for Disordered Systems, Phystech School of Applied Mathematics and Computer Science, Moscow Institute of Physics and Technology, Institutsky Lane 9, Dolgoprudny, Moscow Region, 141700, Russia}

\date{\today}
\begin{abstract}
The 3D digital reconstruction of multi-scale, heterogeneous porous media is traditionally limited by the resolution boundaries and artifacts of X-ray computed tomography (XCT). This paper introduces a physically consistent 2D to 3D pore space synthesis framework that supersedes conventional geometric shuffling with dynamic Newtonian gravitational deposition using a high-performance rigid-body mechanics engine. To honor strict physical contact mechanics, statistical 2D targets, and morphological constraints, we develop a two-stage dynamic packing pipeline that combines realistic 3D grain shapes with a virtual sub-particle insertion engine. The framework is validated against synthetic and natural core benchmarks. First, we show that our stereological integral equation systematically converts size distributions; bypassing this inversion step induces a systematic 50\% deviation in permeability predictions despite identical total porosities, independently altering flow channel topology. Second, temporary virtual spheres serve as architectural placeholders to stabilize large non-equilibrium dissolution voids, ensuring the reconstructed model satisfies target flow metrics within the prescribed tolerance. Finally, when applied to a fine-grained, weakly consolidated sandstone, the workflow exposes and suppresses systematic XCT artifacts, including phantom internal porosity. Ensembles of two-point correlation functions demonstrate that the hydrodynamically and structurally synthesized digital twins quantitatively exhibit closer structural proximity to the uncorrupted rock geometry than the reference XCT dataset itself. This methodology establishes a scalable foundation for high-fidelity multi-phase rock reconstruction, with potential extensions toward complex amorphous cement phases.
\end{abstract}


\maketitle 

\section{Introduction}
\label{sec:intro}

The realistic reconstruction and predictive modeling of complex, heterogeneous porous and granular media remain a grand challenge across materials science, geophysics, and chemical engineering, with direct implications for predicting transport properties, mechanical behavior, and microstructural evolution~\cite{mollon20143d,ju20143d}. 
Historically, the digital rock physics (DRP) revolution has relied heavily on non-destructive X-ray computed microtomography (XCT) to capture internal geometries. 
However, modern investigations of multi-scale, highly heterogeneous, and fine-grained materials hit a fundamental barrier known as the ``curse of resolution-to-field-of-view ratio''. 
For an extensive range of advanced applications, conventional benchtop XCT scanners provide insufficient spatial resolution (typically limited to $\sim 1~\mu\text{m}$), leaving critical sub-micron pore structures, micro-porous cementing phases, and complex grain-to-grain contact zones completely unresolved.

This limitation is particularly acute in several highly relevant domains:
\begin{enumerate}
    \item Unconsolidated and Weakly Cemented Fine-Grained Sandstones: Such fragile systems are prone to mechanical degradation during coring and laboratory testing. 
    Their flow behavior is dominated by sub-micron grain distributions and clay matrices that escape standard XCT visualization~\cite{kulygin2024pore,munawar2018petrographic}.
    \item Ore Deposits and Leaching Fronts: The investigation of acid-induced or water-driven mineral dissolution reveals complex, reactive micro-porosity networks. 
    Tracking how the dissolution of specific mineral phases alters local structural topology and global transport requires sub-micron structural fidelity~\cite{gouze2011xray,wang2024evolution}.
    \item Post-Experimental Dissolution and Matrix Alteration: Materials subjected to hydrodynamic or chemical weathering experience local mineral leaching, transforming stable grain arrangements into highly unstable, poorly sorted granular aggregates with shifting contact networks~\cite{hu2007coupled,gao2025impact}.
\end{enumerate}

To overcome the physical boundaries of XCT, alternative workflows leverage two-dimensional (2D) high-resolution imaging modalities, such as scanning electron microscopy (SEM), focused ion beam SEM (FIB-SEM), and optical microscopy of thin sections. 
These techniques offer cheap, fast, and sub-micron or even nanometer-scale resolution over representative areas.

\begin{figure}[ht]
\centering
\includegraphics[width=0.95\linewidth]{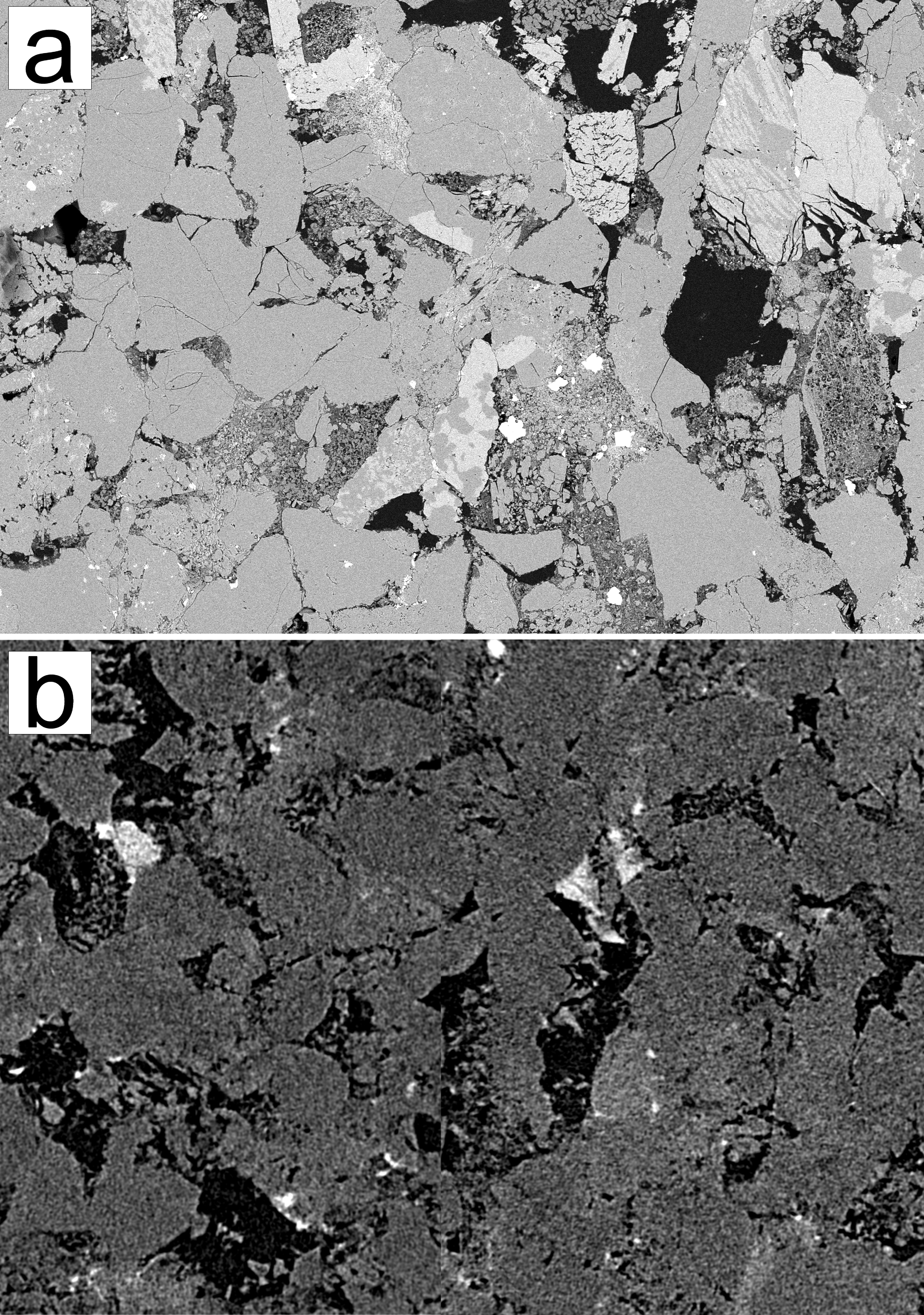}
\caption{
    Motivation for 2D to 3D reconstruction based on XCT limitations.
    (a) High-resolution 2D SEM image: reveals a highly heterogeneous sandstone framework with quartz, carbonate cement, kaolinite aggregates, and altered, leached feldspars hosting complex intragranular micro-porosity within empty interstices (permeability ~50 mD).
    (b) Corresponding 3D XCT slice: demonstrates severe phase blurring due to nearly identical X-ray attenuation coefficients of quartz and feldspars. 
    The scan fails to segment individual grain contacts or classify mineral phases, while partial volume effects fabricate phantom internal micro-pores and artificially choke critical flow pathways.
}
\label{fig: motivation}
\end{figure}

The fundamental necessity of shifting from 3D XCT to high-resolution 2D SEM data for highly heterogeneous, multi-phase systems is vividly demonstrated by a signature sandstone specimen analyzed in this work (Fig.~\ref{fig: motivation}a). 
This poly-mineralic, porous rock matrix exhibits a highly complex clastic framework composed of quartz grains, variable states of altered feldspars, rock fragments, mica flakes, kaolinite aggregates, and carbonate cement. 
Crucially, the pore space is dominated by a severe multi-scale hierarchy, comprising both large intergranular networks and intricate intragranular micro-porosity hosted within porous feldspar remnants formed via intense natural leaching. 
Furthermore, the material distribution is highly non-uniform, separating into tightly packed grain clusters and heavily macro-porous regions, yielding a macroscale absolute permeability of approximately 50 mD.

When subjected to conventional XCT imaging (Fig.~\ref{fig: motivation}b), the fine-grained fractions, micro-porous kaolinite gels, and diffuse boundaries of the leached feldspars escape accurate 3D voxelization due to severe partial volume effects. 
Crucially, since the X-ray attenuation coefficients of quartz and various feldspar phases are nearly identical, standard XCT scans completely fail to segment individual grain contacts, let alone perform an accurate mineralogical classification of the solid matrix. 
The tomographic data inevitably blurs these sharp multi-phase transitions, either artificially closing critical fluid pathways or fabricating phantom voids, thereby failing to capture the true hydrodynamic behavior of the medium. 
Consequently, the scientific focus has shifted toward developing robust stochastic and geometric algorithms capable of reconstructing representative, artifact-free 3D volumes from uncorrupted, high-resolution 2D data pools.

\begin{figure*}[th]
\centering
\includegraphics[width=0.92\linewidth]{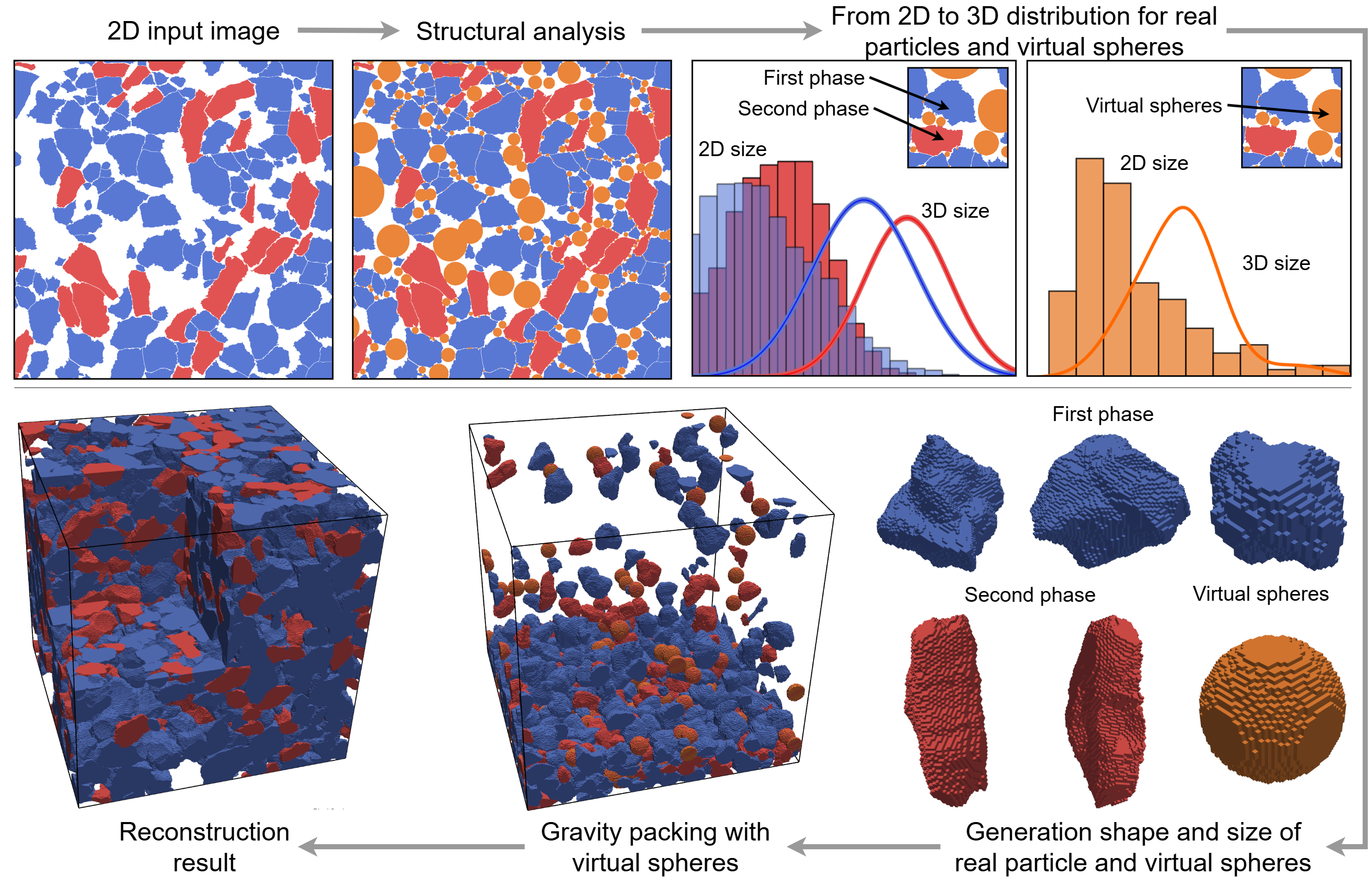}
\caption{Overall workflow for reconstructing granular samples. The process involves, first, structural analysis of the input image, followed by the computation of 2D and 3D size distributions of the structural elements. Next, generation of 3D particle shapes, followed by the gravitational settling of these particles into a specific volume with porosity control. The final reconstruction is obtained from the sedimentation results after removal of virtual spheres.}
\label{fig: pipeline}
\end{figure*}

A granular medium reconstruction process can be naturally decomposed into two fundamental subtasks: the generation of individual grains with prescribed morphological characteristics and their spatial arrangement within the reconstructed domain. 
The latter, often referred to as the packing or placement problem, is particularly challenging due to the need to satisfy geometric constraints (non-overlap, contact conditions) while reproducing target statistical descriptors such as grain size distribution, porosity, and coordination number.

Among the geometric packing strategies, the advancing front algorithm offers a deterministic approach to constructing dense packings of particles with varying shapes~\cite{feng2003filling,morales2016contributions,xia2023minkowski,xia2025efficient}. 
In this method, particles are sequentially added to the system, each placed in contact with one or more previously positioned neighbors. 
Although effective for spherical and simple-shaped particles, the method faces significant computational challenges when extended to arbitrarily shaped grains, often requiring multisphere approximations where complex grains are represented as clusters of overlapping spheres. 
Geometric optimization approaches provide another perspective~\cite{han2005sphere,li2026dense}; for instance, the big-disk-first algorithm leverages Voronoi diagrams to identify optimal placement locations, sorting particles by decreasing size to place them sequentially into the most compact available voids~\cite{ryu2020voropack}. 
Similarly, Voronoi tessellations can be used as a scaffold where cell geometries are iteratively adjusted via an inverse Monte Carlo procedure~\cite{xu2009topological,mollon2012fourier,wang2022novel}. 
While these methods effectively reproduce global statistical descriptors, they frequently fail to produce physically realistic contact networks: grains may remain isolated without true mechanical contacts, or their shapes become overly influenced by the underlying Voronoi geometry, compromising the realism of the resulting microstructure.

An alternative class of methods relies on physically motivated deposition processes. 
The Steepest Descent Ballistic Deposition (SDBD) model provides an efficient framework for simulating the settling of particles under gravity, treating particle motion as a sequence of discrete events (contacts, rolling, pivoting) corresponding to local minima in the gravitational potential~\cite{visscher1972random,jullien2000computer,topic2016steepest}. 
For more detailed physical realism, the Discrete Element Method (DEM) offers a robust alternative by explicitly solving Newton's equations of motion for each particle, including translational and rotational dynamics~\cite{cundall1979discrete,zhu2008discrete,xie2026rock}, often incorporating cohesive interactions like the Johnson-Kendall-Roberts (JKR) contact model~\cite{johnson1971surface,deng2013dynamic}. 
However, DEM simulations reveal non-intuitive, highly complex structural features such as ``cellular'' packings stabilized by cohesion, which can be computationally prohibitive when handling multi-million particle systems with highly irregular shapes. Process-based reconstruction methods complement purely statistical or image-based approaches. In the DEM-based workflow of \cite{wang2024process}, grains are reconstructed from segmented CT images using spherical-harmonic descriptors and packed as irregular clumps to reproduce the thermo-mechanical evolution of pore space under elevated temperature and stress. This formulation emphasizes compaction, thermal damage, and the resulting degradation of pore-throat connectivity and permeability. By contrast, the present method is designed for reconstruction from limited 2D information, where the 3D structure is inferred from 2D statistics and refined with virtual spheres to match the target pore fraction and morphology.

In our previous benchmark study \cite{kulygin2024pore}, we demonstrated the power of a hybrid 3D reconstruction approach for loosely consolidated media. 
That workflow successfully combined analytical grain shape generation from 2D slices with statistical phase retrieval techniques to insert amorphous clay matrices. 
While that framework proved powerful enough to expose and correct subtle laboratory measurement errors (such as the clogging of protective mesh screens), it relied on purely geometric optimization rules during the grain placement stage. 
Overlapping grains were sequentially rearranged, scaled, or rotated based on statistical targets until non-overlap conditions were met. 
Despite its statistical accuracy, such geometric ``shuffling'' lacks physical grounding, failing to capture the inherent stable structures governed by true gravitational deposition and contact mechanics.

To address these fundamental limitations and establish a rigorous, physically consistent paradigm for multi-phase media synthesis, we initiate a comprehensive series of papers dedicated to the high-fidelity reconstruction of complex granular and amorphous structures from 2D data. This paper establishes a standalone, comprehensive methodology focused on replacing traditional heuristic optimization with a physically motivated gravitational deposition framework, operationalized via a high-performance rigid-body dynamics engine. Crucially, to reconcile the strict laws of Newtonian mechanics with the multi-scale statistical constraints of the input 2D images, we introduce a two-stage gravitational packing algorithm. The first stage utilizes realistic grain shapes to form a physically stable backbone, while the second stage dynamically co-deposits spherical sub-particles (virtual spheres) to achieve fine-tuned control over the target porosity and pore size distribution without disrupting the established contact networks. The proposed workflow represents a fully resolved, self-contained scientific contribution capable of generating hydrodynamically equivalent 3D digital twins from uncorrupted 2D datasets, independent of any subsequent algorithmic variations. 

While the modular architecture of this framework naturally allows for future modular enhancements—such as the integration of highly non-convex grain generators or multi-phase amorphous cementing models for complex residual pore voids. The physics-driven regularizations developed and validated in this work provide a complete, autonomous solution to the classical 2D to 3D pore space reconstruction problem.

To demonstrate the versatility, stability, and validity, the developed methodology is applied to three distinct validation cases presented in the results section of this work.

Case 1: Synthetic Sample for Size Distribution Validation, the transition between 2D and 3D particle size distributions, explicitly confirming the mathematical accuracy of our derived integral equation used for stereological matrix transformation.

Case 2: Synthetic Sample for Packing Control in Large Voids. This case tests the limits of porosity regulation within media characterized by significant voids and high porosities, demonstrating how our virtual particle deposition method (corrector spheres) successfully targets and adjusts the structural parameters of the pore space.

Case 3: Reconstruction of a Real Fine-Grained Sandstone Sample. This case applies the full hybrid workflow to a real rock sample for which both high-resolution SEM and benchmark XCT data are available, providing a quantitative comparison of final structural descriptors and absolute hydrodynamic permeabilities to confirm the real-world performance of the method.

\section{Methods}

In this section, we describe the reconstruction workflow used to generate a three-dimensional granular microstructure from a two-dimensional image of the material. The input data may consist of segmented images such as SEM, CT, or QEMSCAN maps, in which the pore phase and the individual mineral phases are identified. The workflow is organized into four sequential stages, and the overall pipeline is summarized in Figure \ref{fig: pipeline}. First, we analyze the 2D geometry of the grains and pore space and extract grain masks and pore descriptors from the image (Section ``Morphological Characterization of Solid and Void Domain'' \ref{pore_analysis}). Second, we infer the 3D pore-size distribution from 2D statistics by solving a stereological inversion problem (Section ``Stereological Inversion and 2D to 3D Particle Size Distribution Transformation'' \ref{distribution}). Third, we reconstruct individual grains from orthogonal 2D masks and generate solid particles consistent with the observed morphology and virtual spheres (Section ``Three-Dimensional Grain Shape Reconstruction from 2D Masks'' \ref{grains_generation}). Finally, we pack the reconstructed grains under gravity in a two-stage procedure with a porosity correction, where temporary virtual spheres are introduced to compensate for porosity and pore-size deficits and are subsequently removed, yielding a dense 3D sample whose grain morphology, pore statistics, solid fractions, and phase proportions match the target 2D structure (Section ``Dual-Stage Gravitational Packing Framework for High-Fidelity Porosity Regulation'' \ref{gravity_packing}).

\subsection{Morphological Characterization of Solid and Void Domain} \label{pore_analysis}

\begin{figure}[th]
\centering
\includegraphics[width=0.92\linewidth]{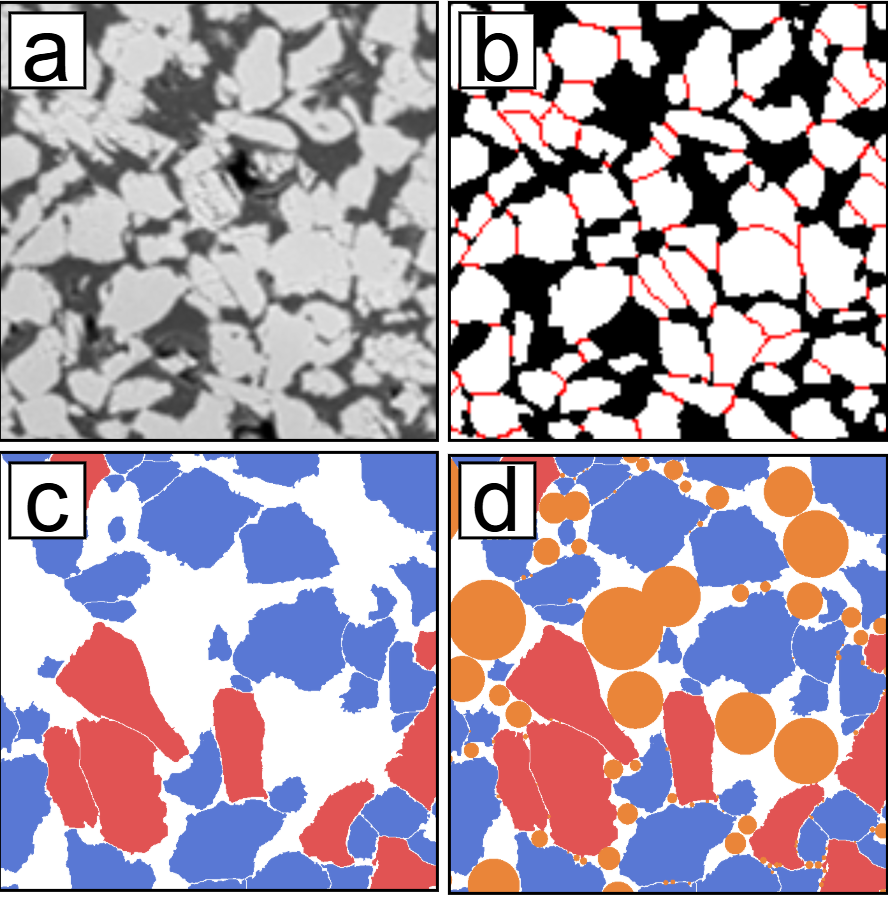}
\caption{Grain mask extraction and pore space analysis. (a) SEM image; (b) Binarized SEM image with separation into individual masks (red lines); (c) Binarized image with two grain phases (shown in red and blue); (d)  Binarized image with inscribed circles of maximum radius (in orange).}
\label{fig: mask}
\end{figure}

This section describes the preprocessing of the 2D input image required to construct a statistically representative 3D granular packing. The primary objective is to obtain, from a segmented image, two key parameters: the 2D grain-size statistics of the solid phases and the pore-size statistics of the void space, and to extract 2D masks of individual grains of the solid phases. These grain masks are then used to reconstruct the grain shape in 3D, similar to these masks. Grain-size statistics are then used to initialize the size of the generated grains. Pore-size statistics are then used to control the subsequent gravitational sedimentation stage. 

The first step is the extraction of individual grain masks for each mineral phase. We apply a segmentation procedure based on Morse theory \cite{zubov2022pore, gostick2019porespy}, which separates touching grains and preserves the contour of each object. Figure \ref{fig: mask} (b) shows the binary grain image together with the partition into individual masks separated by red lines. Detected masks are labeled by mineral phase (grains type) according to the segmentation labels or by subsequent feature-based classification, as illustrated in Figure \ref{fig: mask}(c). For each detected mask, the equivalent 2D radius is computed as $r = \sqrt[2]{{S_{2D}}/{\pi}}$, where $S_{2D}$ is the area of the mask. This yields the empirical grain radius distribution $f^{P_i}_{2D}(r)$, where ${P_i}$ denotes the phase number.

\begin{figure}[th]
\centering
\includegraphics[width=0.94\linewidth]{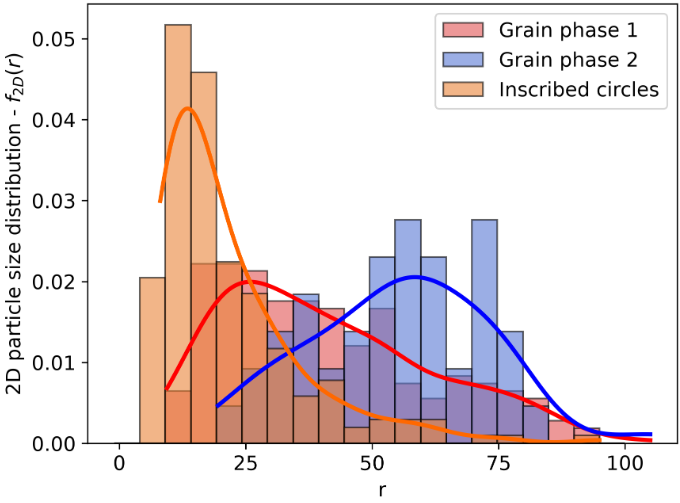}
\caption{Visualization of particle size distributions $f_{2D}(r)$ for two granular phases (in red and blue) and pore inscribed circles (in orange).}
\label{fig: 2d_histograms}
\end{figure}

The second stage concerns the pore space. The pore phase is separated from the grain phases, and the void domain is decomposed into individual pore regions using the same topological segmentation approach \cite{zubov2022pore, gostick2019porespy}. For each pore, the maximal inscribed circle is then constructed from the distance transform, as illustrated in Figure \ref{fig: mask}(d). In this Figure, circles denote the largest inscribed circles within each pore. Each pore is therefore characterized by a representative radius, and the complete set of these radii defines the 2D pore size distribution $f^{sp}_{2D}(r)$, see example in Figure \ref{fig: 2d_histograms}.

\subsection{Stereological Inversion and 2D to 3D Particle Size Distribution Transformation
} \label{distribution}

The segmentation stage yields empirical 2D grain-size distributions $f^{P_i}_{2D}(r)$, where $r$ denotes the equivalent radius associated with each grain mask. To construct a physically consistent 3D packing, we reconstruct the underlying 3D distribution $f^{P_i}_{3D}(R)$ from $f^{P_i}_{2D}(r)$. This is a stereological inversion problem (Wicksell's corpuscle problem \cite{wicksell1925corpuscle, depriester2019resolution}): a grain is visible in a 2D section only when the section intersects the particle, and therefore the apparent section radius is generally smaller than the true grain radius.

To describe the relationship between the distributions of particle radii in 2D and 3D, a model is used in which each particle is represented as a sphere in 3D and, correspondingly, as a circle in an arbitrary 2D cross‑section of the model particle. Accordingly, the 3D grain radius is defined as $R = \sqrt[3]{{3V_{3D}}/{4\pi}}$, where $V_{3D}$ is the volume of the particle.

\begin{figure}[th]
\centering
\includegraphics[width=0.6\linewidth]{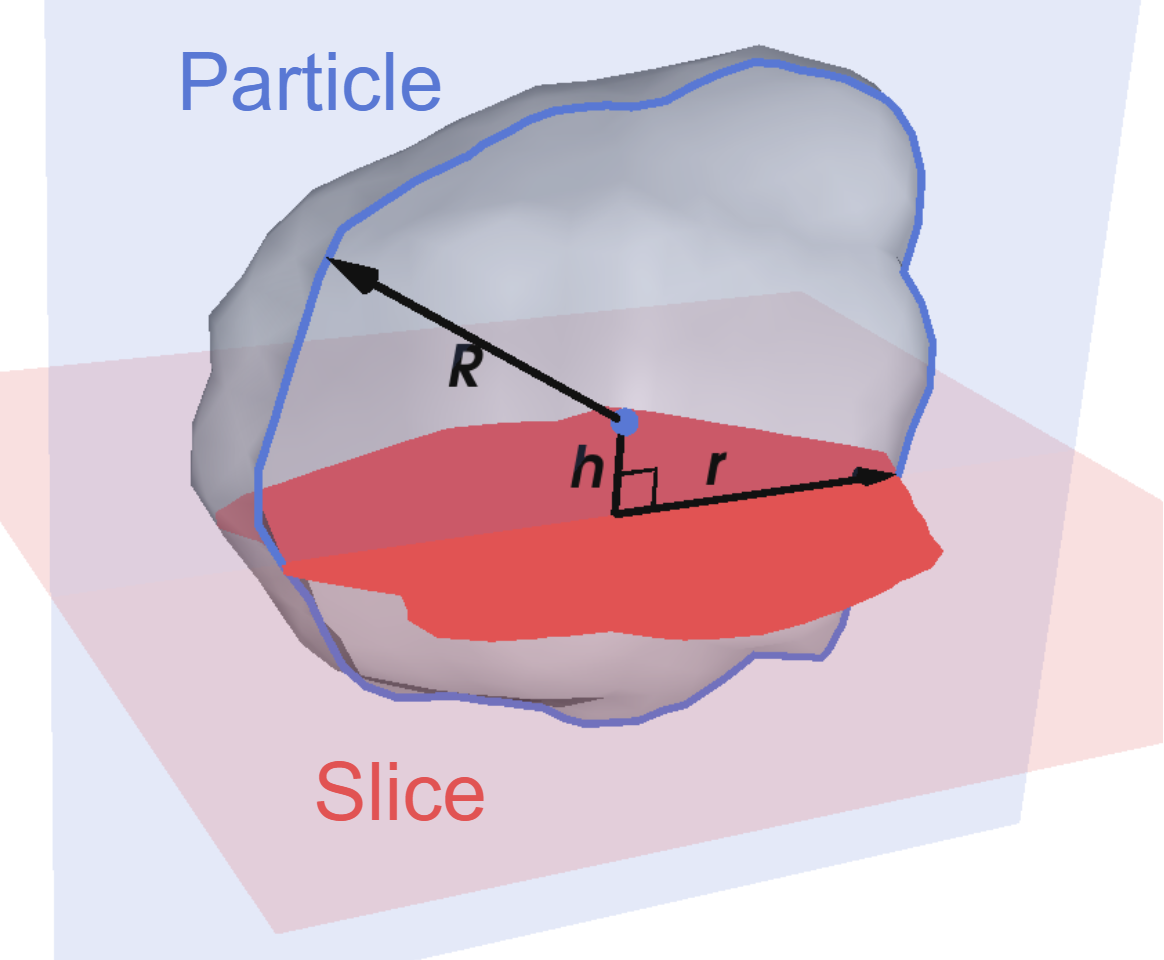}
\caption{Schematic model of a particle intersecting a thin slice: the 3D particle is shown in blue and the 2D slice plane -- in red.}
\label{fig: model}
\end{figure}

Consider the probability that an arbitrary 2D slice contains a particle with radius $r$. Consider a 3D particle with fixed radius $R$, see Figure \ref{fig: model}. The probability density that the particle intersects the selected arbitrary slice is:

\begin{equation}
    f_1 \sim 2R
\end{equation}

If the particle intersects the slice, the radial size in the section depends on the distance $h$ from the center of the particle to the slice plane. Assuming a uniform distribution of $h$ over $[0,R]$, the relation $R^2 = h^2 + r^2$ implies $h\,dh = r\,dr$. Hence the probability density of observing a 2D radius $r$ is:

\begin{equation}
    f_2 \sim \frac{dh}{R} \cdot \frac{1}{dr}= \frac{r}{R\sqrt[2]{R^2 - r^2}}
\end{equation}

Thus, the probability density that a 2D slice shows a particle with radius $r$ given that it originates from a 3D particle of radius $R$ is:

\begin{equation}
    f(r \mid R) = f_1 \cdot f_2
    \sim
    \begin{cases}
        \displaystyle \frac{r}{\sqrt[2]{R^2 - r^2}}, & r < R \\
        \,\,\,\,\,\,\,\,\,\,\,\,0, &  r \ge R
    \end{cases}
    \label{eq: main_eq}
\end{equation}

Taking into account the 3D distribution $f_{3D}(R)$, the probability density that a particle observed in the 2D slice with radius $r$ originates from a 3D grain of radius $R$ is (where $C$ denotes a constant):

\begin{equation}
    f_{2D}(r) = \int^{\infty}_{r} f(r \mid R)\,f_{3D}(R)\, dR = \cfrac{r}{C} \int^{\infty}_{r} \frac{f_{3D}(R)\,dR}{\sqrt[2]{R^2 - r^2}}
    \label{eq: integr}
\end{equation}

The distribution $f_{2D}(r)$ is computed from the input image and represented as a histogram with $N$ bins and values $w^{2D}=(w_{1}^{2D}, \dots, w_{N}^{2D})$. The bin boundaries $r_0, r_1, \dots, r_N$ are determined based on the maximum grain radius in the 2D slice, $r_N = 2 \cdot \max[r]$, so that the densities $f_{2D}(r)$ and $f_{3D}(r)$ are effectively zero beyond $r_N$. The distribution $f_{3D}(R)$ is then recovered by solving Equation \ref{eq: integr} and represented as a histogram with the same number of bins and unknown weights $w^{3D}= (w_{1}^{3D}, \dots, w_{N}^{3D})$, with same bins position. Each histogram bin $w_{i}^{2D}$ of the $f_{2D}(r)$ distribution is computed as follows:

\begin{equation}
    w_{i}^{2D} = \int_{r_{i-1}}^{r_i} \cfrac{f_{2D}(r)\,dr}{r_i - r_{i-1}} =  \int_{r_{i-1}}^{r_i} \cfrac{r\,dr}{r_i - r_{i-1}} \cdot \cfrac{I}{C}
\end{equation}

\begin{equation}
I=\int^{\infty}_{r} \frac{f_{3D}(R)\,dR}{\sqrt[2]{R^2 - r^2}}
\end{equation}

Several simplifications are employed. The lower limit of the inner integration is varied from $r$ to the minimum of $r_i$ and $r_{i+1}$, and the integral is then decomposed into several sub integrals:

\begin{equation}
    I = \int^{\infty}_{r_{i-1}} \frac{f_{3D}(R)\,dR}{\sqrt[2]{R^2 - r^2}} 
    =\sum^{N}_{j = i} \int^{r_{j}}_{r_{j-1}} \frac{f_{3D}(R)\,dR}{\sqrt[2]{R^2 - r^2}}
\end{equation}

When $R$ varies across the bin width from $r_{i-1}$ to $r_i$, we assume that $f(R)$ is constant and equal to the value at the midpoint of the bin, i.e., $f(R) = f(r_{m,j})$ with $r_{m,j} = (r_j + r_{j-1})/2$:

\begin{equation}
    I= \cfrac{1}{2} \sum^{N}_{j = i} f_{3D}(r_{m,j}) \cdot\ln \Biggl[\cfrac{(R + \sqrt[2]{R^2 - r^2})}{(R - \sqrt[2]{R^2 - r^2})}\Biggl]^{R=r_j}_{R=r_{j-1}}
\end{equation}

Therefore, the expressions for $w_{i}^{2D}$ are as follows:

\begin{equation}
    \begin{split}
        w_{i}^{2D} &= \sum^{N}_{j = i} \int_{r_{i-1}}^{r_i} \cfrac{f_{3D} (r_{m,j})\,dr}{r_i - r_{i-1}} \cdot \cfrac{r}{2C} \cdot \\
        &\cdot \ln \Biggl[\cfrac{(R + \sqrt[2]{R^2 - r^2})}{(R - \sqrt[2]{R^2 - r^2})}\Biggl]^{R=r_j}_{R=r_{j-1}}
    \end{split}
\end{equation}

The approximation is used with $r_{m,i} = (r_i + r_{i-1})/2$ and $dr =  r_i - r_{i-1}  = \mathrm{const}$, where the value of the function is treated as constant within each bin:

\begin{multline}
    w_{i}^{2D} = \sum^{N}_{j = i} w_{j}^{3D} \cdot \cfrac{r_{m,i}}{2C} \cdot \ln \Biggl[\cfrac{(R + \sqrt[2]{R^2 - r^2_{m,i}})}{(R - \sqrt[2]{R^2 -  r^2_{m,i} })}\Biggl]^{R=r_j}_{R=r_{j-1}}
    \label{eq: 1}
\end{multline}

This leads to a linear system: 

\begin{equation}
    \mathbf{A} w^{3D} = w^{2D}
    \label{eq: 2}
\end{equation}

where the matrix $\mathbf{A} = (a_{ij})$ is given by:

\begin{equation}
    a_{ij} =
    \begin{cases}
                 r_{m,i} \cdot \ln \Biggl[\cfrac{(R + \sqrt[2]{R^2 - r^2_{m,i}})}{(R - \sqrt[2]{R^2 -  r^2_{m,i} })}\Biggl]^{R=r_j}_{R=r_{j-1}}, & i \le j \\
        \,\,\,\,\,\,\,\,\,\,\,\,0, &  i  > j
    \end{cases}
    \label{eq: a}
\end{equation}

The linear system in Equation 11 is solved for the unknown vector $(w_1^{3D},\dots,w_N^{3D})$ using a non-negative least squares (NNLS) solver with Tikhonov regularization to mitigate the ill-posed nature of the Volterra integral inversion and eliminate non-physical negative weights. The solution is then normalized to satisfy the probability constraint. The same procedure is applied independently for each granular phase and for the sizes of the inscribed circles distribution. The reconstructed 3D grain-size distribution is then used to generate particles whose radii are consistent with the original microstructure and to initialize the subsequent gravitational packing stage.

\subsection{Three-Dimensional Grain Shape Reconstruction from 2D Masks} \label{grains_generation}

To reconstruct the geometry of individual grains, we follow the approach described in \cite{mollon2013generating, kulygin2024pore}, where a 3D particle is recovered from three orthogonal 2D masks. The central idea is to use the observed contour information as geometric constraints for the particle surface so that the final grain reproduces the measured sections in the principal directions while remaining smooth and physically plausible in 3D.

\begin{figure}[th]
\centering
\includegraphics[width=0.94\linewidth]{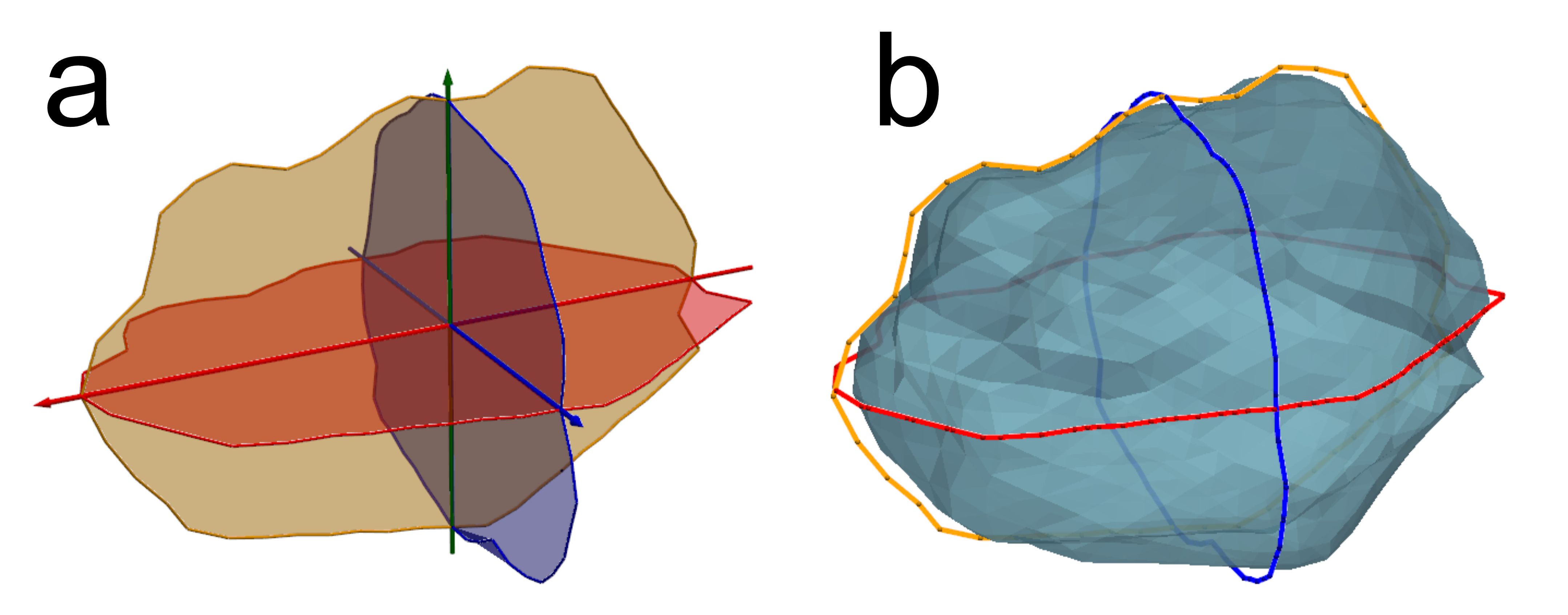}
\caption{Grain reconstruction from three orthogonal masks. (a) Three mutually aligned orthogonal masks of grains; (b) Reconstructed grain surface based on these masks.}
\label{fig: grain}
\end{figure}

Since the input imaging provides only planar 2D profiles, the three orthogonal masks required for a single 3D grain reconstruction are sampled stochastically from the statistical ensemble of extracted 2D grain masks within the corresponding size bin, assuming macroscopic structural isotropy of the clastic framework. For each grain, three masks are arranged in space with coincident centers and mutually orthogonal orientations. Their contours are locally compressed or stretched to align them with the $X$, $Y$, and $Z$ coordinate axes, as shown in Figure \ref{fig: grain}(a). The reconstruction proceeds by splitting the two vertical masks into halves and generating four intermediate rotational bodies from these half-contours. A horizontal mask is then imposed on these intermediate solids so that every cross-section contains a scaled version of the observed horizontal profile. The final grain is obtained by interpolating between these four shapes; as a result, the reconstructed grain preserves the original masks in three orthogonal central sections, as shown in Figure \ref{fig: grain}(b).

Once the particle geometry is reconstructed, its characteristic size is rescaled according to the recovered 3D size distribution from Section \ref{distribution}. The same procedure is applied to the pore-space circles extracted in Section \ref{pore_analysis}: for these void elements, spheres are generated with radii inferred from the reconstructed $f^{sp}_{3D}(r)$. This step ensures that all solid components are statistically consistent with the initial 2D microstructure before the gravitational packing stage begins.

\subsection{Dual-Stage Gravitational Packing Framework for High-Fidelity Porosity Regulation} \label{gravity_packing}

\begin{figure*}
\centering
\includegraphics[width=0.92\linewidth]{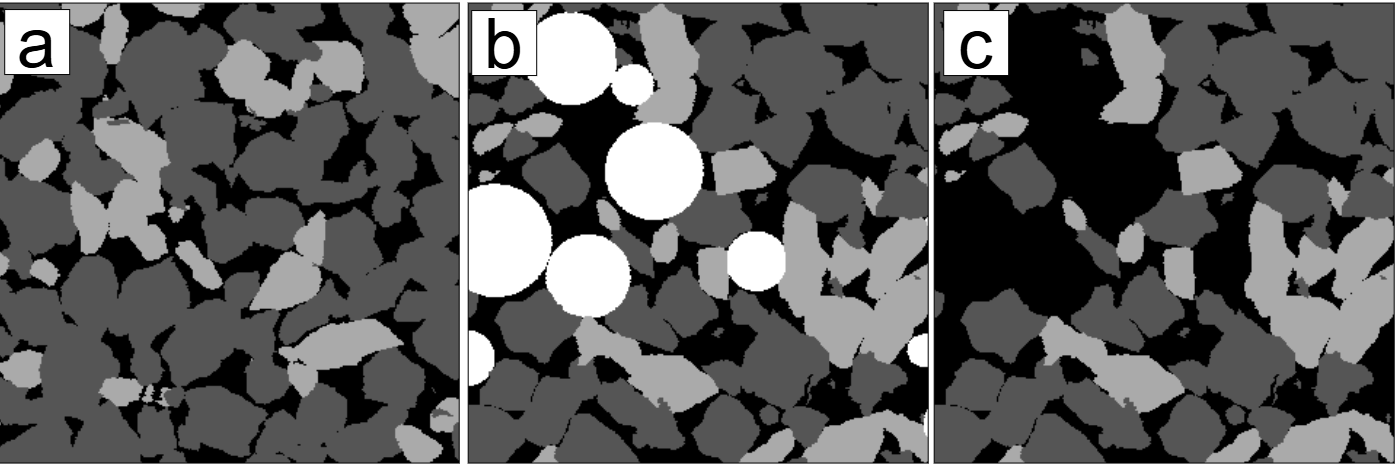}
\caption{Two-stage packing for porosity and pore-size control. (a) 2D slice after the first packing (reconstructed grains only); (b) slice after the second packing (grains and virtual spheres); (c) final slice after removal of virtual spheres.}
\label{fig: two_packing}
\end{figure*}

To reconstruct the sample, two successive gravitational packing stages are used. The first packing stage produces a baseline reconstruction from the generated grains and gives the reference pore structure generated by this reconstructed grains. The second stage introduces a virtual spherical phase whose size distribution is chosen to compensate for the mismatch between the baseline pore statistics and the target statistics extracted from the 2D input image. After the virtual spheres are removed, the result preserves the reconstructed grain morphology while providing a pore space whose porosity and pore-size distribution are consistent with the target 2D structure.

\textbf{Physical modeling.} 
Physics engines such as PhysX can be used for a wide range of simulation tasks involving rigid bodies, collisions, contact constraints, and constrained virtual environments \cite{macklin2013position, erez2015simulation, maciel2009using, al2019particula}. The motion of the grains is simulated using the NVIDIA PhysX engine, which models rigid-body dynamics according to Newtonian mechanics. Each grain is represented as a dynamic rigid body with mass, inertia tensor, and linear and angular velocities. At each time step, the engine integrates acceleration under gravity, updates velocities, and advances positions. The time integration is discretized in a manner similar to a semi-implicit Euler scheme, which improves numerical stability while preserving the efficiency of the simulation.

Collision detection is performed in two stages. During the broad phase, bounding-volume tests are used to discard non-interacting particle pairs; during the narrow phase, contact constraints are evaluated using a persistent contact manifold and GJK-based geometric checks \cite{gilbert1988fast}. Contact resolution is performed by the impulse method: normal impulses eliminate interpenetration, while tangential impulses enforce Coulomb friction \cite{baraff1994fast}. The restitution coefficient controls the fraction of the normal relative velocity retained after impact and is prescribed through the material parameters. 

The same simulation settings were used for all three datasets considered in this study. The basic parameters that govern the reconstruction process and influence the resulting sample morphology are summarized in Appendix \ref{Appendix}. These are the essential baseline settings required to reproduce the gravitational packing procedure.

\textbf{Two-stage packing process.} The target properties of the reconstructed sample are the pore fraction $\phi_{pore}^{target}$, the fractions of the mineral phases $(\phi_1^{target},\dots,\phi_N^{target})$, the grain shapes, grain and pore size distribution. The reconstruction procedure therefore consists of two gravitational packing steps. In the first step, reconstructed grains are packed without virtual spheres -- first packing. The resulting porosity and pore-size statistics of first packing are compared with the corresponding target values extracted from the input 2D image. In the second step, virtual spherical particles and reconstructed grains are packed together -- second packing, so that the pore geometry and porosity of second packing match the target morphology and porosity.

\begin{figure}
\centering
\includegraphics[width=0.95\linewidth]{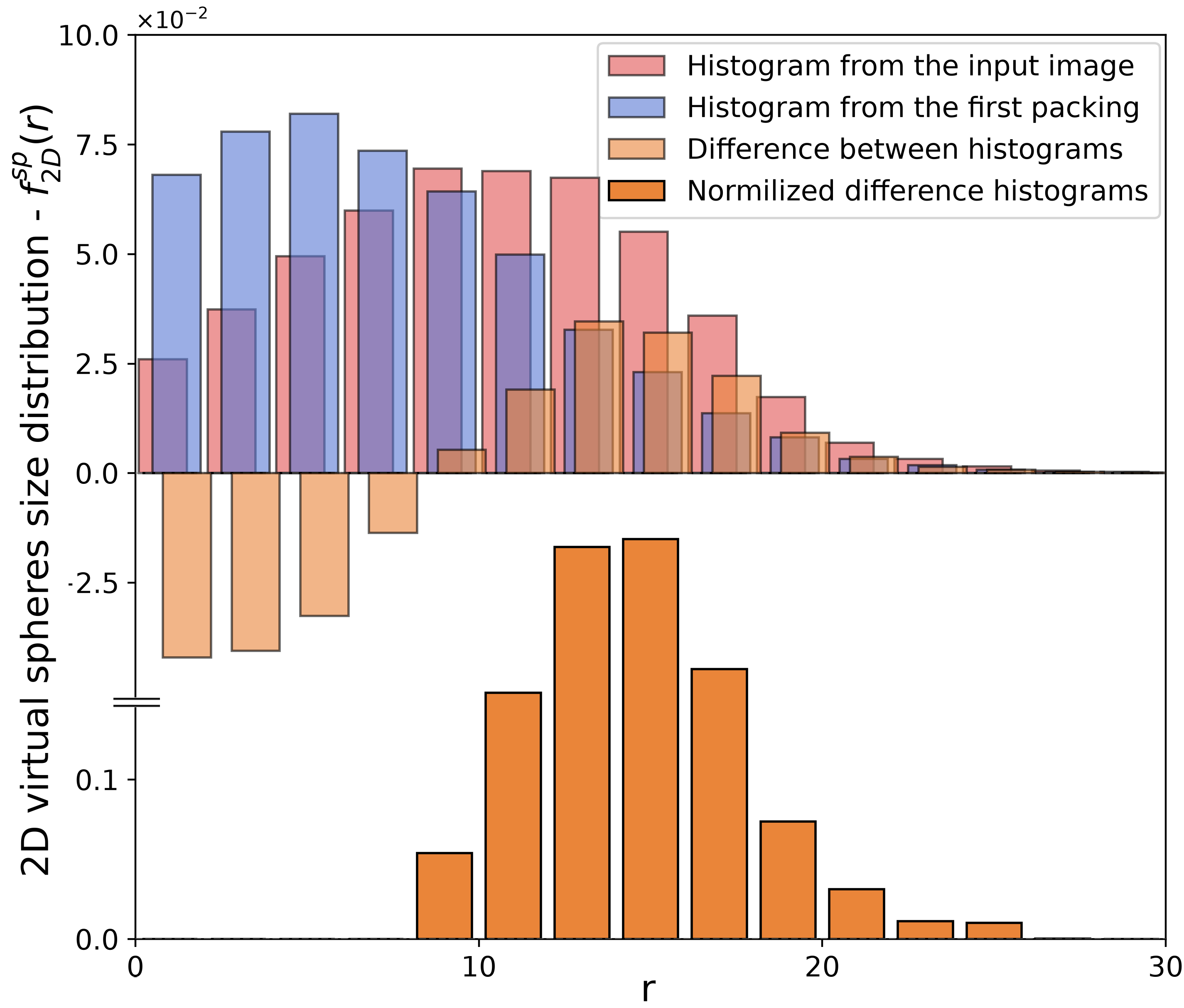}
\caption{Distributions of virtual 3D spheres used in the second packing. Histograms of the virtual-sphere size distribution computed from the input image and from slices of the first packing (in red and blue). The difference (in light orange) defines the virtual-sphere size distribution (in orange) used in the second packing.}
\label{fig: delta_hists}
\end{figure}

The result of the second packing is illustrated in Figure \ref{fig: two_packing}(a). The first packing provides an estimate of the pore fraction $\phi_{pore}^{(1)}$ and a first approximation of the pore-size statistics that occurs during gravitational sedimentation. To evaluate these quantities, representative slices are selected from the reconstructed volume and analyzed with the same circle extraction procedure described in Section \ref{pore_analysis}. For all slices, the largest inscribed circles are computed and converted into a 2D pore-size histogram $f_{2D}^{sp,(1)}(r)$, see red histogram in Figure \ref{fig: delta_hists}. This histogram is then transformed into a 3D estimate $f_{3D}^{sp,(1)}(r)$ using the inversion procedure described in Section \ref{distribution}. The resulting distribution characterizes the size of the pore elements naturally produced by the first packing stage.

The target pore-size distribution is obtained from the original image in the same way (as described in Sections \ref{pore_analysis} and \ref{distribution}): the largest inscribed circles in the segmented pore phase are computed, and the corresponding 3D histogram $f_{3D}^{sp}(r)$ is recovered, see blue histogram in Figure \ref{fig: delta_hists}. The difference between the baseline and target distributions is then used to construct the size distribution of virtual spherical particles:

\begin{equation}
    f_{3D}^{sp,(2)}(r) = f_{3D}^{sp}(r) - f_{3D}^{sp,(1)}(r)
    \label{eq: delta_hist}
\end{equation}

where negative values are set to zero and the distribution is subsequently normalized. The result is a  distribution of virtual sphere radii that are added to the packing to compensate for the deficit of pore space elements in the first reconstruction. The corresponding histogram is shown in orange in Figure \ref{fig: delta_hists}.

To determine the number of virtual particles to be packed, let $p_{sp}$ be the probability that the next packed particle is a virtual sphere. In the first packing, the solid volume fraction is $\phi_{solid}^{(1)} = 1 - \phi_{pore}^{(1)}$, whereas in the second packing the solid fraction excluding the virtual spheres is $\phi_{solid}^{(2)} = 1 - \phi_{pore}^{target}$. Since the total solid volume remains unchanged between the two packings, it follows that:

\begin{equation}
    \phi_{solid}^{(1)} = \phi_{solid}^{(2)} + \phi_{sp}
\end{equation}

where $\phi_{sp}$ is the total volume fraction of the virtual spherical particles. Denote $V_{sp}$ as the mean volume of a virtual sphere, computed from $f_{3D}^{sp,(2)}(r)$, and let $V_p$ denote the mean volume of a reconstructed grain from all phases calculated from $f_{3D}^{P_i}(r)$. The ratio between the non-spherical volume fraction and spherical solid volume fraction contributions is:

\begin{equation}
    \frac{\phi_{solid}^{(2)}}{\phi_{sp}} = \frac{(1-p_{sp})V_p}{p_{sp}V_{sp}}
\end{equation}

Therefore, the probability is given by:

\begin{equation}
    p_{sp} = \frac{1}{1+k}, \qquad
    k = \frac{1-\phi_{pore}^{target}}{\phi_{pore}^{target} - \phi_{pore}^{(1)}} \cdot \frac{V_{sp}}{V_p}
\end{equation}

The result of the second packing is illustrated in Figure \ref{fig: two_packing}(b). After the virtual spheres are removed, the final reconstruction is shown in Figure \ref{fig: two_packing}(c). This step yields a sample whose pore geometry and porosity are close to the target statistics extracted from the 2D image. Owing to the stochastic nature of gravitational deposition, the reconstructed porosity may differ from the target value by a small amount, typically below $0.03$ in absolute value. Such a discrepancy is expected in a statistically generated packing.  In addition, the porosity of a thin 2D section can vary locally from one specimen window to another, even when the same material is considered. When a more precise match to the target porosity is required, a final small adjustment is applied to the reconstructed packing.

This correction is deliberately minimal. The positions of the packed grains are scaled isotropically about the center of the reconstructed sample, while their individual geometry is preserved. The adjustment is limited to the small difference needed to compensate for the residual mismatch in porosity; therefore, the grains only penetrate one another marginally and the number of contacts remains practically unchanged. The contact areas are modified only slightly, and the pore connectivity and the topological organization of the pore space are preserved. In this sense, the final correction acts as a weak porosity-tuning operation rather than a new packing process, and it does not alter the essential morphology of the reconstructed medium.

\textbf{Control of the multimineral volume fractions.} During both packing processes, the number of grains of each granular phase is determined by their sampling probability. Let $M$ denote the number of grain phases and $p_i$ denote the probability that a grain from phase $i$ is selected during packing, with $\sum_{i=1}^{M} p_i = 1$. From the segmented 2D image, the target volume fractions of the granular phases relative to the total solid volume are computed as $\phi_1,\dots,\phi_M$, with $\sum_{i=1}^{M}\phi_i=1$. For each phase, we compute the mean grain volume $V_i$ from the 3D particle size distribution $f^{P_i}_{3D}(r)$. Then the volume fraction of phase $i$ in the reconstructed sample is:

\begin{equation}
    \phi_i' = \phi_i = \frac{p_iV_i}{\sum_{j=1}^{M} p_jV_j} \qquad i=1,\dots,M.
\end{equation}

Probabilities $p_i$ are computed via solution of the linear system:

\begin{equation}
    \mathbf{B}\mathbf{p} = \mathbf{c}
\end{equation}

where $\mathbf{p} = (p_1,\dots,p_{M-1})^{\top}$ and the matrix $\mathbf{B}=(b_{ij})$ of size $(M-1)\times(M-1)$:

\begin{equation}
    b_{ij} = \phi_i V_j - \delta_{ij}V_j
\end{equation}

and the vector $\mathbf{c} = (c_i)$ of size $M-1$:

\begin{equation}
    c_i = -\phi_iV_M
\end{equation}

The solution is then extended by setting $p_M = 1 - \sum_{i=1}^{M-1}p_i$ and renormalized to enforce $\sum_{i=1}^{M}p_i=1$. During the packing procedure, grains from phase $i$ are sampled with probability $p_i$, which yields a reconstructed solid fraction consistent with the input 2D image.

\section{Reconstruction Quality Metrics}

To rigorously evaluate the fidelity and predictive capacity of the developed hybrid reconstruction framework, two distinct categories of descriptors are utilized: a statistical microstructural descriptor and a macroscale transport property. 
These metrics ensure that the generated 3D models are equivalent to the target media both geometrically and hydrodynamically~\cite{cherkasov2021adaptive}.

\subsection{Two-Point Correlation Function \texorpdfstring{$S_2(r)$}{S2(r)}}

To assess the spatial, morphological, and statistical equivalence of the reconstructed grain matrix to the original media, we employ the two-point probability function $S_2(r)$~\cite{torquato2002random,yeong1998reconstructing,postnicov2023surface,postnicov2024evaluation}. 
While simple volumetric parameters like porosity provide only a first-order scalar approximation of the pore space, they completely lack information regarding the spatial architecture, internal connectivity, and layout of the phases. 
In contrast, the two-point correlation function serves as a fundamental, mathematically rigorous statistical descriptor for disordered and heterogeneous materials~\cite{torquato2002random}. 
It encapsulates crucial microstructural features, including characteristic grain sizes, specific surface areas, cluster distribution lengths, and the overall degree of structural disorder within the medium~\cite{torquato2002random,cherkasov2021adaptive}. 
Utilizing $S_2(r)$ for quality evaluation is highly meaningful, as it acts as a sensitive geometric "fingerprint" capable of detecting subtle spatial anisotropies and structural deviations that remain completely invisible to standard visual or scalar comparisons.

For a binary medium, the solid phase is defined by an indicator function $I(\mathbf{x})$, where $I(\mathbf{x}) = 1$ if $\mathbf{x}$ belongs to the solid grain phase and $I(\mathbf{x}) = 0$ otherwise. 
The two-point correlation function represents the probability that two randomly selected points separated by a distance vector $\mathbf{r}$ both fall within the solid phase. 
It is mathematically expressed as:

\begin{equation}
S_{2}(\mathbf{x}_{1},\mathbf{x}_{2})=\langle I(\mathbf{x}_{1})I(\mathbf{x}_{2})\rangle
\end{equation}

where $\langle \dots \rangle$ denotes ensemble averaging. 
Under the assumption of statistical homogeneity and isotropy, which generally holds true for representative volumes of sedimentary rock matrices, the directional vector simplifies to a scalar distance $r = \vert{}\mathbf{x}_1 - \mathbf{x}_2\vert{}$.

The boundary behaviors of $S_2(r)$ provide immediate physical insights into the system's global properties. 
At zero distance ($r=0$), the function yields the absolute solid phase fraction:

\begin{equation}
S_{2}(0)=\phi _{s}=1-\phi 
\end{equation}

As the distance approaches infinity ($r \rightarrow \infty$), the spatial correlation between the two points naturally vanishes, and the function asymptotically levels off to the square of the solid phase fraction~\cite{jiao2007modeling}:

\begin{equation}
\lim_{r \to \infty } S_{2}(r) = \phi _{s}^{2}
\end{equation}

The intermediate slope and the rate of decay of the $S_2(r)$ curve between these two boundaries are directly governed by the structural characteristics of the media. 
The initial derivative of the function at zero distance is strictly related to the specific surface area of the pore-grain interface, while any oscillatory behavior or damping lengths at larger values of $r$ explicitly reveal the characteristic correlation lengths and structural periodicities of the grain packing.
Comparing the $S_2(r)$ curves of the original images against the statistical cloud of the reconstructed models therefore offers an absolute metric of morphological validity, ensuring that the dynamic physical packing honors the complete spatial configuration of the natural sample.

\subsection{Pore-Network Extraction and Absolute Permeability}

While statistical microstructural descriptors validate spatial and morphological similarity, the definitive test for high-fidelity digital replicas is their capacity to accurately reproduce macroscale transport properties. 
In this work, single-phase fluid flow is simulated using a topologically consistent pore-network modeling (PNM) approach~\cite{blunt2001flow,blunt2002detailed,xiong2016review}.

Pore-network modeling has emerged as a globally recognized and widely implemented framework across a vast spectrum of scientific and engineering disciplines to evaluate hydrodynamic, mechanical, and thermal properties of disordered media. 
Its applications extend far beyond traditional earth sciences, ranging from the structural analysis of amorphous molecular polymer matrices~\cite{ananev2026structure} and the investigation of nano-confined flow and phase behavior during pore-to-ensemble scale transitions~\cite{nesterova2025bridging}, to the characterization of complex soil macro-structures~\cite{tolstygin2026soil} and advanced digital rock physics~\cite{zhang2024hybrid,shi2025pore}.

The strategic choice of the PNM approach over direct grid-based simulation techniques—which solve transport equations directly on the voxelized geometry (e.g., direct Navier-Stokes solvers or the Lattice Boltzmann Method)—is fundamentally dictated by the multi-scale, highly heterogeneous nature of our target materials. 
To reconstruct a truly representative elemental volume (REV) of fine-grained or unconsolidated media with multi-scale architecture, the simulation domains must be exceptionally large~\cite{zubov2024rev}, routinely yielding final digital models on the order of billions of voxels ($10^9$ voxels).
Simulating fluid dynamics on grids of such massive proportions presents a formidable, often prohibitive computational challenge for direct numerical methods. 
In contrast, the pore-network approximation serves as a highly robust, computationally elegant, and scalable alternative. 
It compresses the massive voxel data into a graph while rigorously retaining the core physics, making transport calculations on billion-voxel datasets entirely feasible on accessible hardware.

The pore space of the voxelized 3D models is converted into a graph of interconnected pores (nodes) and throats (edges) using an extraction algorithm based on discrete Morse theory and persistent homology, a method that mathematically guarantees the conservation of the original pore space topology~\cite{zubov2022pore}.
For each individual pore-to-pore connection, the local hydraulic conductance $g_{ij}$ is parameterized using the classical circle-triangle-square model~\cite{mason1991capillary,oren1998extending,patzek2001shape,valvatne2004predictive} to accurately capture the impact of cross-sectional shape factors.
Single-phase fluid flow through the network is then governed by mass conservation enforced at each individual pore node:

\begin{equation}
    \sum_{j} q_{ij} = 0, \quad q_{ij} = \frac{g_{ij}}{L_{ij}} \Delta P_{ij}
\end{equation}

where $q_{ij}$ is the volumetric flow rate between pores $i$ and $j$, $L_{ij}$ is the characteristic length, and $\Delta P_{ij}$ is the local pressure drop. 
Solving the resulting system of linear algebraic equations under specified pressure boundary conditions yields the total volumetric flow rate $q$. 
Finally, the macroscopic absolute permeability $K$ is computed via Darcy's law:

\begin{figure*}
\begin{minipage}[t]{0.9\linewidth}
\centering
\includegraphics[width=0.94\linewidth]{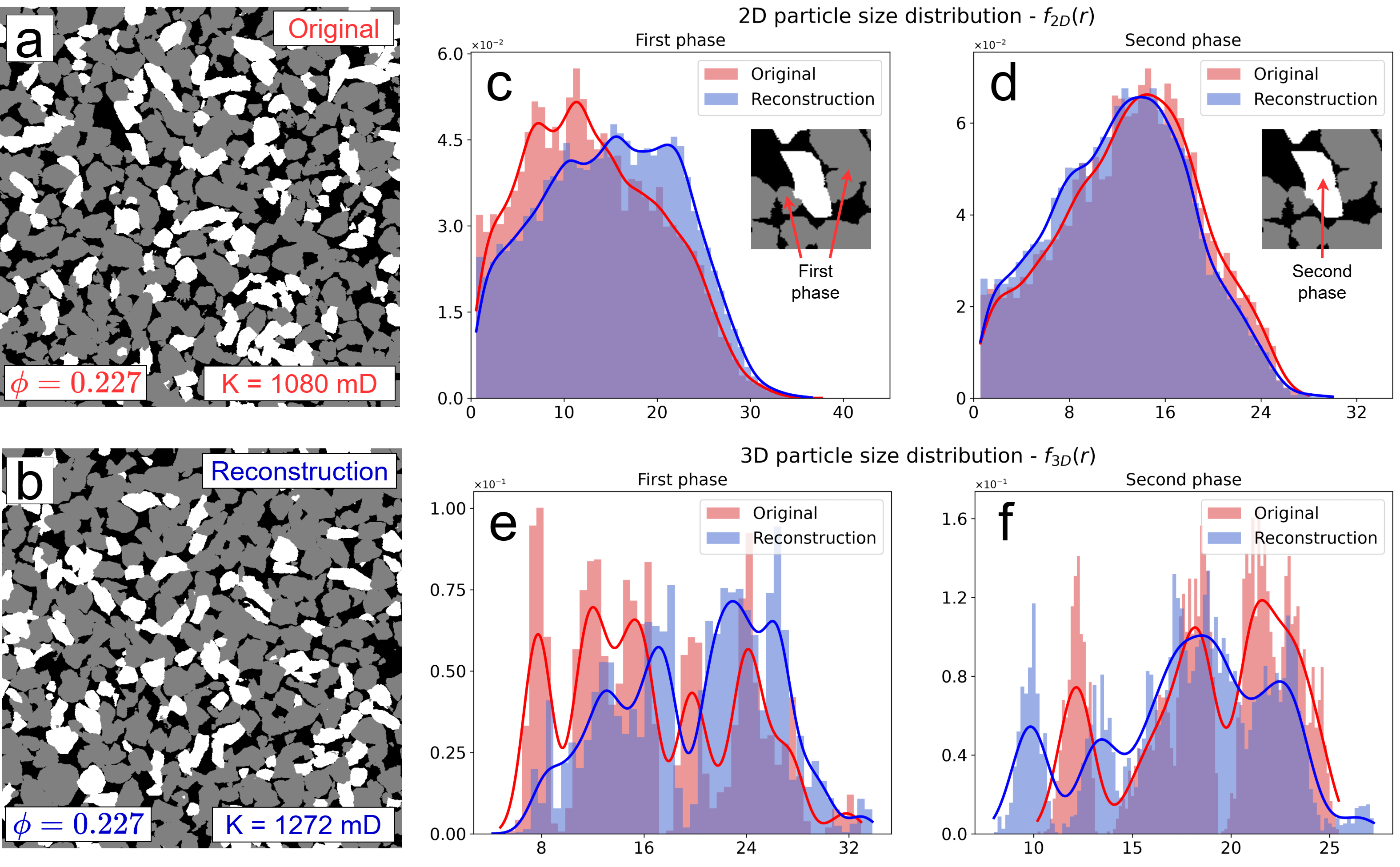}
\caption{Sample 1: with recalculation 2D to 3D distribution (Reconstruction 1). (a) 2D slice of the original synthetic data; (b) 2D slice of the reconstructed data; (c, d) Histograms of 2D radii distributions for the first and second granular phases (original vs. reconstructed); (e, f) Histograms of 3D radii distributions for the first and second granular phases (original vs. reconstructed).}

\label{fig: sample_1_good}
\end{minipage}

\hfill
\vspace{\columnsep}

\begin{minipage}[h]{0.9\linewidth}
\centering
\includegraphics[width=0.94\linewidth]{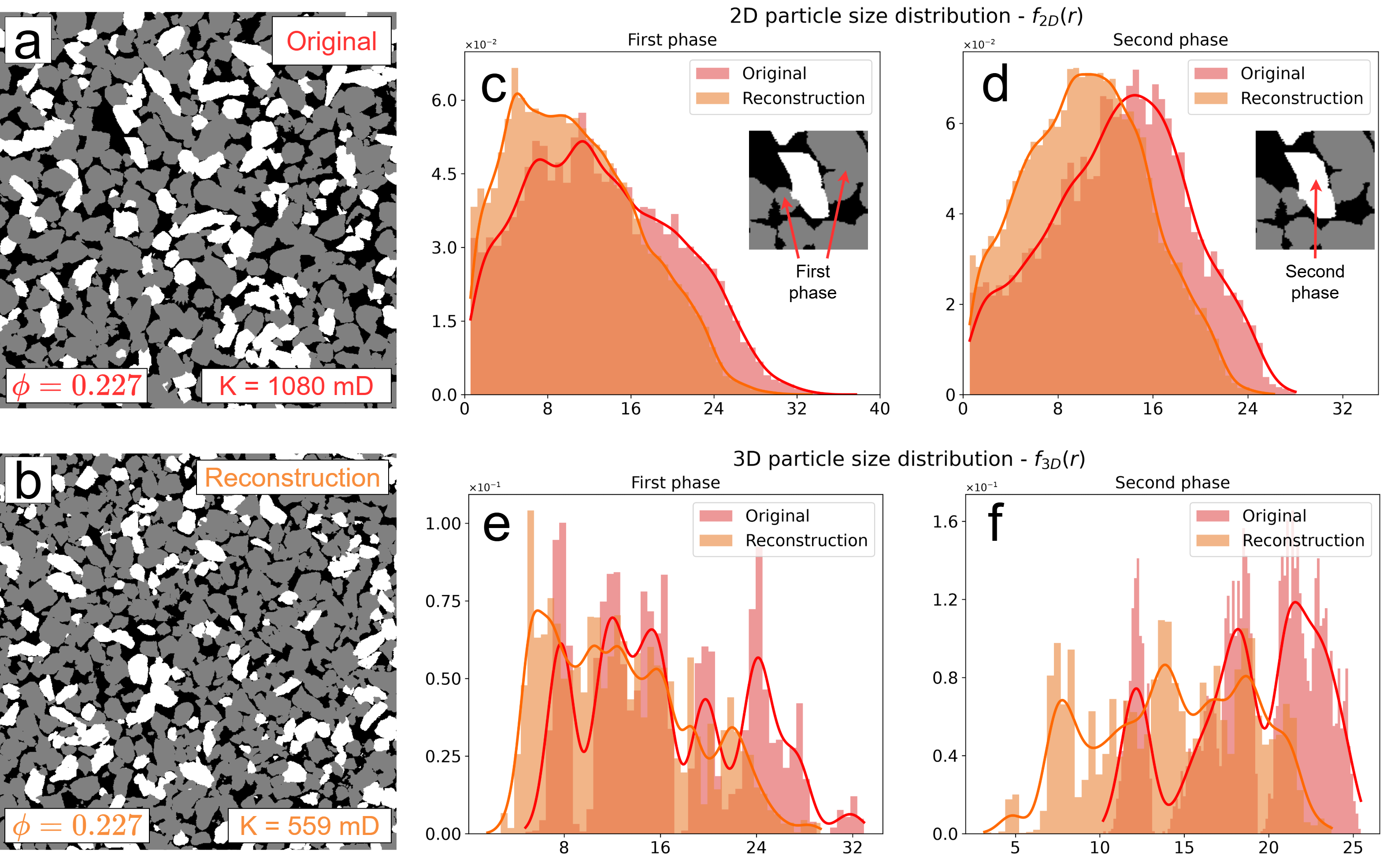}
\caption{Sample 1: without recalculation 2D to 3D distribution (Reconstruction 2). (a) 2D slice of the original synthetic data; (b) 2D slice of the reconstructed data; (c, d) Histograms of 2D radii distributions for the first and second granular phases (original vs. reconstructed); (e, f) Histograms of 3D radii distributions for the first and second granular phases (original vs. reconstructed).}
\label{fig: sample_1_bad}
\end{minipage}
\end{figure*}

\begin{equation}
    K = \frac{\mu q L}{A \Delta P}
\end{equation}

where $\mu$ is the fluid viscosity, $L$ is the total domain length, $A$ is the cross-sectional area, and $\Delta P$ is the global pressure drop across the sample.

\section{Results}

\subsection{Quantitative Validation of Stereological Matrix Inversion on Synthetic Media} 
\label{sub:case1}

The primary objective of this validation case is to rigorously evaluate the performance of our mathematical framework in resolving the classical stereological problem, specifically, the transition from observable 2D slice parameters to the true 3D particle size distribution $f_{3D}(R)$ via the derived integral kernel inversion (Eqs.~\ref{eq: integr} and \ref{eq: 1}). 
In digital rock physics workflows based on 2D images (such as SEM or thin sections), neglecting stereological effects introduces massive structural errors. 
Because a random 2D plane intersects spheres at arbitrary distances from their centers, the apparent 2D radii within $f_{2D}(r)$ are inherently underestimated, shifting the empirical curves toward artificially small values even for perfectly uniform grains.

To validate our matrix inversion approach, a benchmark synthetic granular medium was generated with a strictly predefined, multi-phase 3D grain size distribution $f_{3D}(R)$. A random 2D cross-sectional slice was then extracted from this reference system. 
Individual 2D grain masks were isolated using the Morse-theory segmentation pipeline, and their equivalent 2D radii were computed based on mask areas to establish the input $f_{2D}(r)$ distribution histograms. The synthetic volume for this Sample consists of $1000^3$ voxels.

The structural and hydrodynamic consequences of this validation are presented in Fig.~\ref{fig: sample_1_good} and Fig.~\ref{fig: sample_1_bad}. 
When the stereological matrix inversion is fully active (Sample 1: with recalculation, Fig.~\ref{fig: sample_1_good}), the back-calculated 3D particle size distribution demonstrates an exceptional, seamless agreement with the original benchmark curves for both granular phases. 
The matching of the particle size distribution functions confirms that the analytical integral kernel $a_{ij}$ (Eq.~\ref{eq: a}) successfully eliminates the artificial ``small-grain shift'' caused by random geometric slicing.

To evaluate how these structural corrections influence macroscale transport properties, we simulated a controlled scenario where the 2D to 3D inversion block was deliberately bypassed (Sample 1: without recalculation, Fig.~\ref{fig: sample_1_bad}). 
In this flawed pipeline, the raw, uncorrected 2D size metrics were used directly as a proxy for 3D grain sizes during the dynamic physical packing stage. 
Crucially, during this reconstruction, the target overall porosity was strictly maintained to match the original benchmark ($\phi = 0.227$). 
Despite this volumetric constraint, using artificially shrunken grains forced the NVIDIA PhysX engine to generate a heavily distorted grain backbone. 

\begin{figure}[th]
\centering
\includegraphics[width=0.94\linewidth]{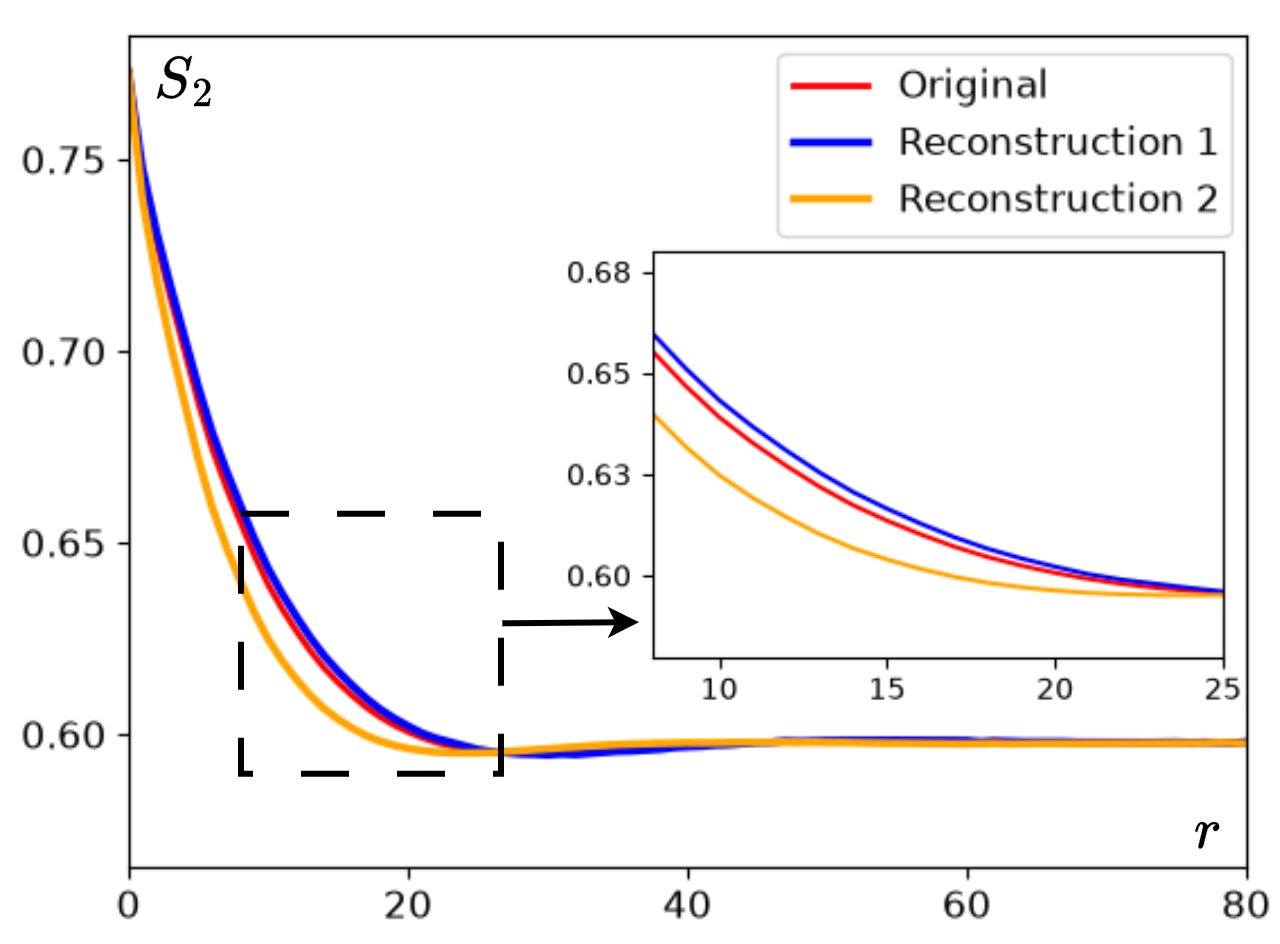}
\caption{Sample 1: two-point correlation function $S_2(r)$. $S_2(r)$ curve of original synthetic data in red, $S_2(r)$ curve of reconstructed data with and without computation 2D to 3D distribution in blue and orange, accordingly.}
\label{fig: sample_1_s2}
\end{figure}

This morphological distortion is quantified by the behavior of the two-point correlation function $S_2(r)$, shown in Fig.~\ref{fig: sample_1_s2}. The $S_2(r)$ curve for the mathematically corrected model (Reconstruction 1) closely aligns with the original data line, demonstrating structural and volumetric consistency within stochastic variances.
Conversely, the uncorrected model's curve (Reconstruction 2) experiences a severe spatial deviation, indicating a fundamental mismatch in phase spatial distribution and a collapse of the effective pore throat architecture.

This structural alteration translates directly into a massive failure of transport property predictions. 
For the true baseline synthetic medium, the absolute permeability calculated via pore-network modeling is $K = 1080~\text{mD}$. 
The model generated with our stereological recalculation block (Reconstruction 1) yields $K = 1272~\text{mD}$, demonstrating a close and acceptable agreement well within the stochastic variance of dynamic packings. 
However, for the uncorrected model (Reconstruction 2), the calculated absolute permeability dropped drastically to $K = 559~\text{mD}$---representing a systematic underestimation of approximately 50\%. 

\begin{figure*}
\begin{minipage}[t]{0.99\linewidth}
\centering
\includegraphics[width=0.94\linewidth]{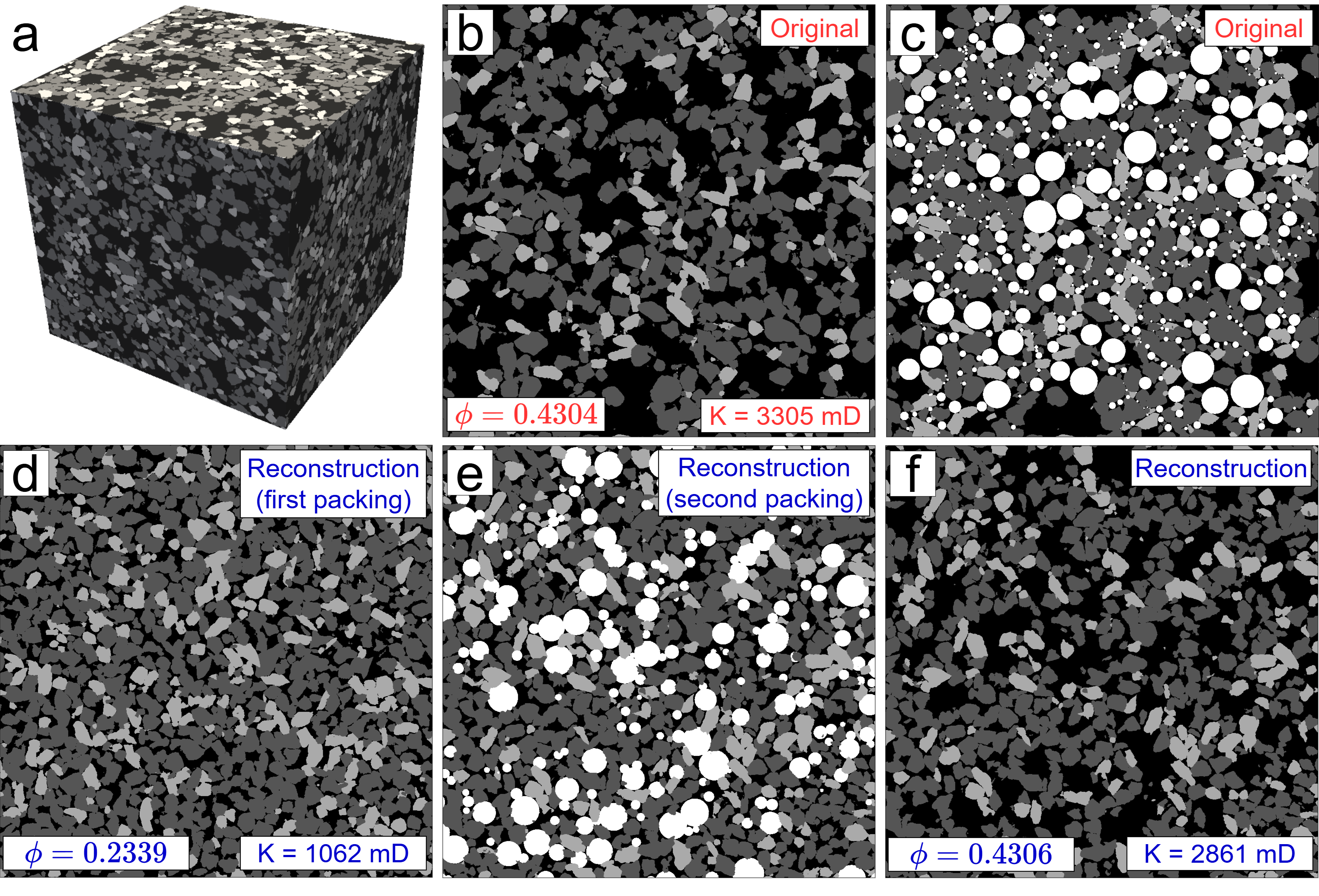}
\caption{Sample 2: comparison of original and reconstructed synthetic data. (a) 3D original synthetic volume; (b) 2D slice of the original synthetic data; (c) 2D slice of the original synthetic data with maximal inscribed circles; (d) 2D slice of reconstructed data after first packing; (e) 2D slice of reconstructed data after second packing with spheres; (f) 2D slice of the final reconstructed data.}
\label{fig: sample_2_result}
\end{minipage}

\hfill
\vspace{\columnsep}

\begin{minipage}[h]{0.99\linewidth}
\centering
\includegraphics[width=0.95\linewidth]{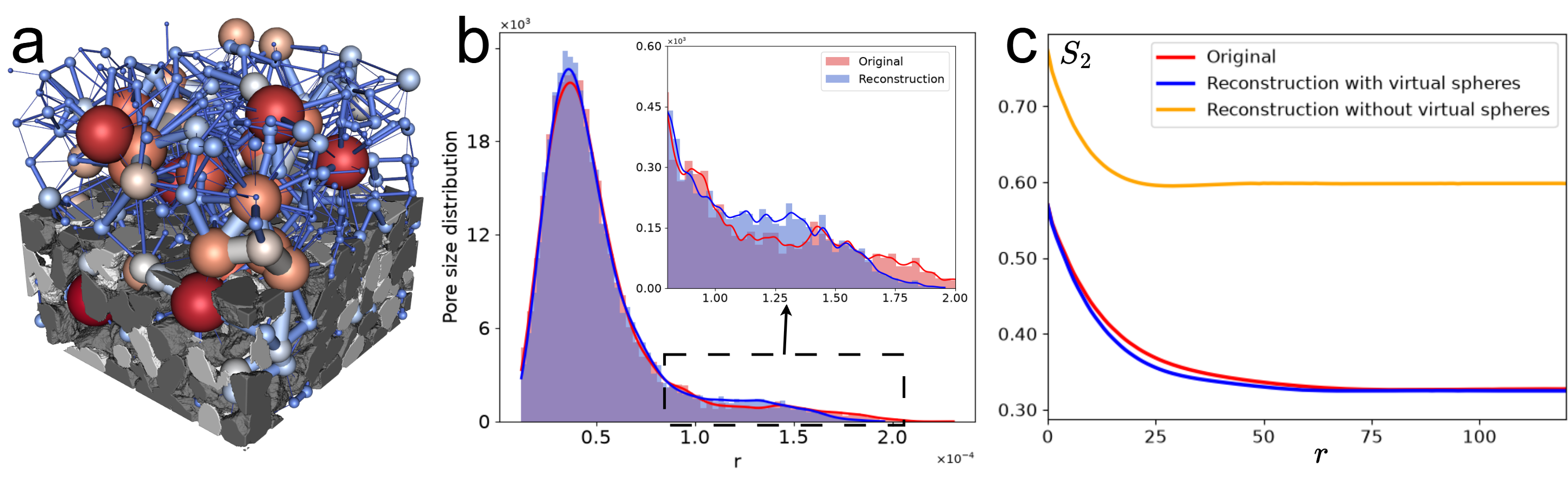}
\caption{Sample 2: structural analysis. (a) Extracted pore-network from voxel data; (b) Comparison of pore network elements size distributions for original and reconstructed data (inset: zoom); (c) Comparison of $S_2(r)$ curves for original data (red), after first packing data (orange), after second packing data (blue).}
\label{fig: sample_2_dist}
\end{minipage}
\end{figure*}

Critically, because the total porosity was kept identical ($\phi = 0.227$) in both reconstructions, this massive drop in permeability is not a trivial consequence of bulk volume reduction or compaction. 
Instead, it explicitly demonstrates that failing to resolve stereological effects chokes the primary hydraulic pathways by altering the pore-to-throat size distribution and connectivity.

\subsection{Porosity Regulation in Heterogeneous Media with Non-Equilibrium Dissolution Voids}
\label{sub:case2}

In the second validation case, we test the operational limits of our virtual particle deposition framework within a highly challenging and non-equilibrium pore void architecture. 
The baseline target medium is characterized by a significantly high overall porosity and a severely heterogeneous pore size distribution dominated by large, interconnected macro-porous voids (Fig.~\ref{fig: sample_2_result} (a), ``Original''). The synthetic benchmark volume for second Sample was selected to $1000^3$ voxels.

From a practical digital rock physics perspective, this synthetic configuration directly models two real-world reservoir scenarios. 
First, it perfectly replicates the structural consequences of experimental mineral dissolution, such as matrix alteration caused by water flooding, reactive acid-induced leaching, or chemical weathering, where specific mineral components are dissolved, leaving behind unstable macro-porous channels. 
Second, this case serves as the foundational intermediate stage for complex multi-phase rock reconstruction. 
In a layered, highly clay-rich or vuggy formation, the dynamic packing framework must first synthesize a stable, open grain skeleton, preserving specific macro-voids that will subsequently be saturated with multi-phase amorphous cementing phases, non-granular clays, or secondary precipitates during the advanced random media synthesis planned for Part 3 of this research cycle.

Standard gravitational packings of complex-shaped grains inherently fail in these scenarios; driven by Newtonian forces, irregular particles naturally settle into their tightest potential minima, causing the large structural voids to collapse as smaller sand or feldspar grains migrate and choke the primary conduits. 
This physical limitation is clearly demonstrated in the first packing stage (Fig.~\ref{fig: sample_2_result} (d), ``Reconstruction (first packing)''), where only the realistic grain shapes are allowed to settle. 
The absence of a structural placeholder forces the grain backbone into an artificially over-compacted state, truncating the total porosity to an unacceptable value of $\phi = 0.2339$ and generating a critically low absolute permeability of $K = 1062~\text{mD}$. 
This structural collapse is quantitatively exposed by the severe spatial deviation of the two-point correlation function $S_2(r)$, as shown in Fig.~\ref{fig: sample_2_dist} (c) (``Reconstruction without virtual spheres'').

To resolve this issue and match the target microstructural descriptors, the algorithm computes the stereological difference between the 2D inscribed circle histograms of the original input image and the slices of the over-compacted first packing (Eq.~\ref{eq: delta_hist}). 
Through the histogram subtraction block, a target size distribution ($f^{sp,(2)}_{3D}(r)$) and a co-deposition probability ($p_{sp}$) for temporary corrector spheres are established.

During the second packing stage (Fig.~\ref{fig: sample_2_result} (e), ``Reconstruction (second packing)''), these virtual corrector spheres are co-deposited alongside the granular phases inside the NVIDIA PhysX engine. 
Acting as temporary structural shields, these spheres mechanically prevent small grains from blocking the macro-pores. 
Once the dynamic settling reaches equilibrium, the temporary spheres are computationally removed, leaving behind the final reconstructed multi-phase matrix (Fig.~\ref{fig: sample_2_result} (f)).

The accuracy of this two-stage porosity control engine is highly stable.
The reconstructed model achieved a final porosity of $\phi = 0.4306$, matching the original target value ($\phi = 0.4304$) with an absolute error of just $0.0002$ ($0.02\%$), well within the specified tolerance. 

More importantly, as demonstrated in Fig.~\ref{fig: sample_2_dist} (c), the ``Reconstruction with virtual spheres'' curve perfectly captures the extended, elongated tail of large pore radii, which represents the preserved dissolution channels. 
This precise structural reconstruction translates directly into hydrodynamic equivalence. 
The absolute permeability of the original benchmark is $K = 3305~\text{mD}$, while our final reconstructed twin yields $K = 2861~\text{mD}$. 
Furthermore, the reconstruction of the microstructural architecture is mathematically evaluated by the two-point correlation function $S_2(r)$ (Fig.~\ref{fig: sample_2_dist} (c)), where the ``Reconstruction with virtual spheres'' curve demonstrates a strong convergence with the original benchmark line. 
This case successfully validates that the insertion of temporary virtual sub-particles provides a mathematically rigorous and physically sound methodology to preserve complex, high-porosity structural heterogeneities independently of grain shape limitations.

\subsection{Reconstruction of a Real Fine-Grained Sandstone Sample}

\begin{figure*}[th]
\centering
\includegraphics[width=0.94\linewidth]{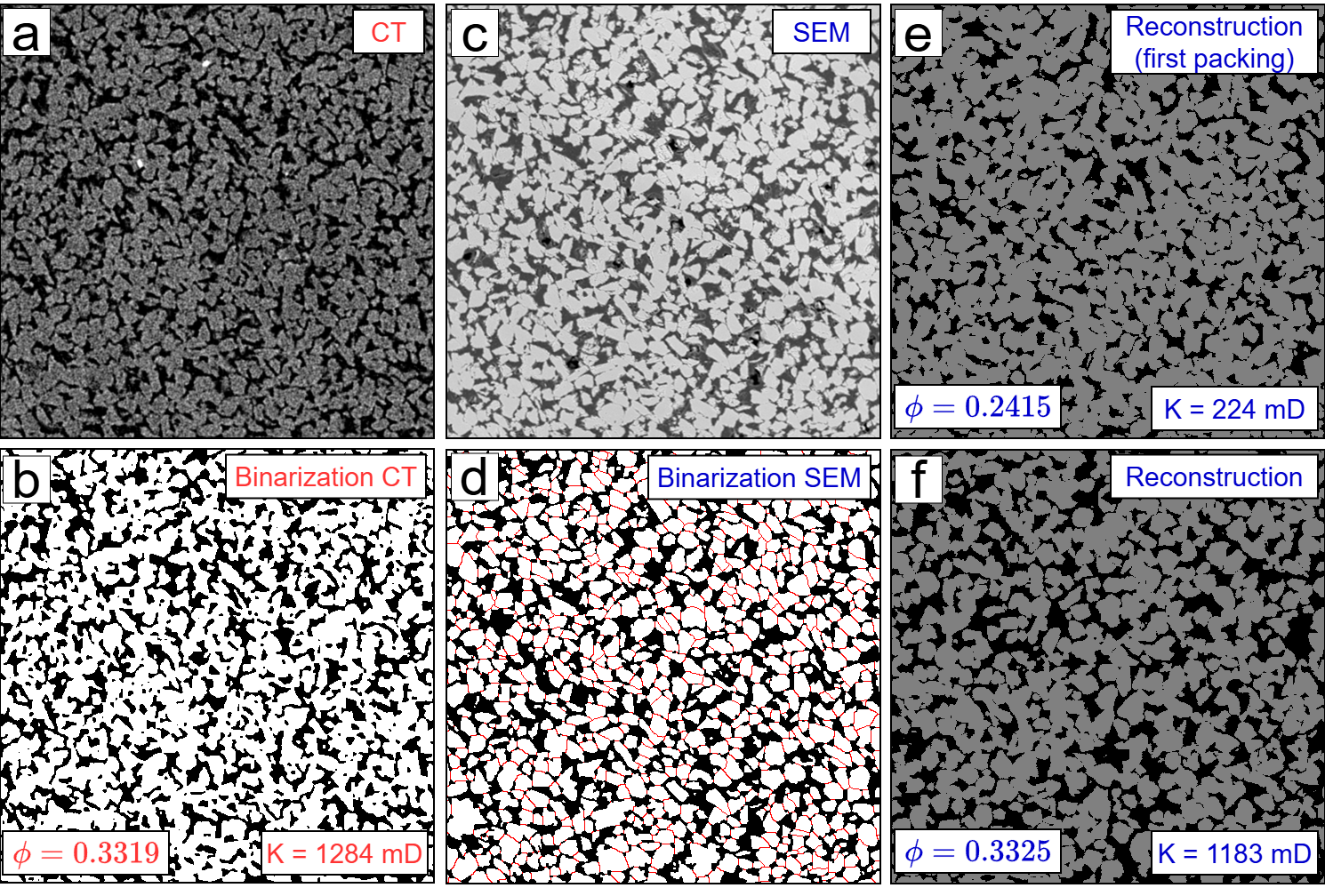}
\caption{Sample 3: comparison of experimental and reconstructed data. (a) 2D slice of CT data; (b) 2D slice of binarized CT slice; (c) SEM image; (d) Binarized SEM image; (e) 2D slice of reconstructed SEM data after first packing; (f) 2D slice of reconstructed SEM data after second packing.}
\label{fig: sample_3}
\end{figure*}

\begin{figure*}[th]
\centering
\includegraphics[width=0.94\linewidth]{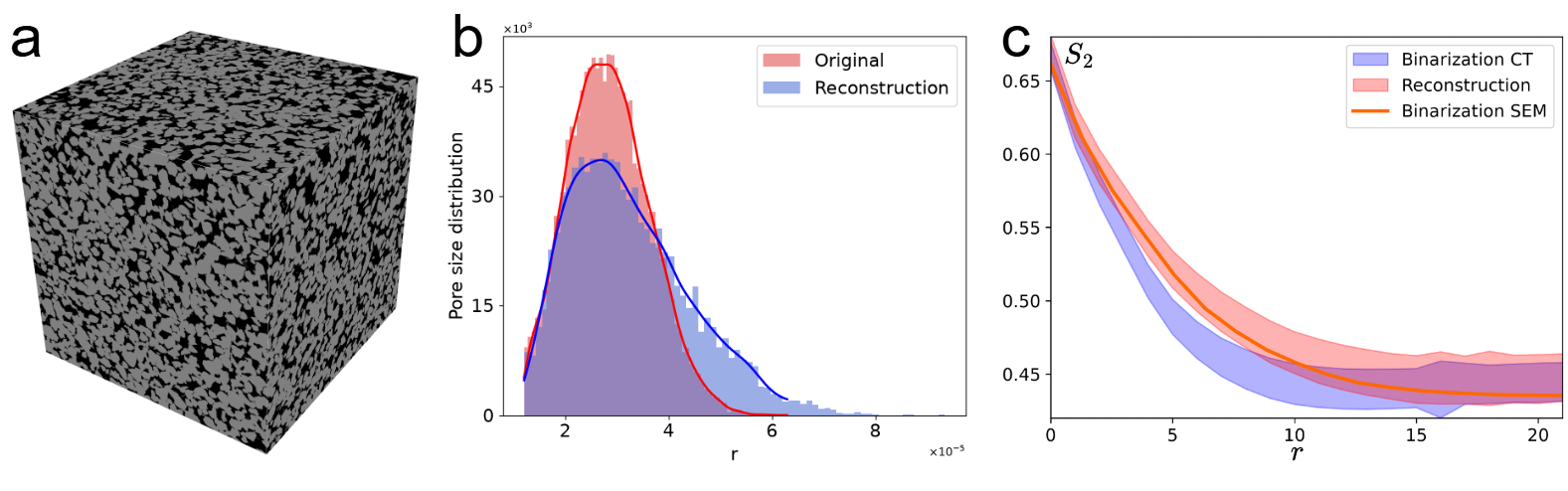}
\caption{Sample 3: structural analysis. (a) 3D reconstructed volume; (b) Comparison of pore network elements size distributions for original and reconstructed data; (c) Comparison of $S_2(r)$ curves for CT slices (blue), reconstructed SEM slices (red), binarized SEM (orange).}
\label{fig: sample_3_dist}
\end{figure*}

The final validation step involves applying the complete workflow to a challenging natural core specimen (Sample 3). 
This sample represents a classical fine-grained, well-sorted, massive porous sandstone characterized by a high degree of uniformity and exceptionally small structural elements. 
Crucially, the internal interstices of this sample are genuinely empty, while its boundary matrix is known to be highly susceptible to localized experimental alterations introduced during physical core sampling.

This specific configuration embodies the exact real-world scenario for which our physics-driven methodology provides the highest utility. 
In digital rock physics of fragile, loosely consolidated formations, a notorious yet frequently overlooked artifact is the mechanical compaction of loose grains at the boundaries of the extracted cylindrical plug during the drilling and trimming stages. 
This process artificially increases the solid phase fraction and densifies the grain configuration near the sample edges. 
By leveraging high-resolution, uncorrupted internal SEM images sampled away from the damaged core margins, the workflow preserves the pristine, non-equilibrium structural state of the rock matrix.

Furthermore, this physical boundary distortion in XCT is compounded by tomographic partial volume artifacts, which falsely identify phantom micro-pores deep inside the solid massive grains (Fig.~\ref{fig: sample_3}(b)). 
Since the sample's interstices are empty, these inner-grain cavities are purely numerical illusions caused by the XCT's inability to cleanly resolve sharp phase boundaries in uniform media with fine-grained components. 
The phantom internal porosity forces an unphysical, systematic shift in the XCT-derived pore size distribution toward the ultra-small radius domain (Fig.~\ref{fig: sample_3_dist}(b)).

The execution of our two-stage dynamic packing pipeline successfully circumvents both tomographic limitations. The reconstructed sandstone volume for Sample 3 was set to $488^3$ voxels. 
Reconstructing the grain skeleton solely from high-resolution 2D masks without porosity control leads to a total gravitational collapse of the loose sand structure ($\phi = 0.2415$, $K = 224~\text{mD}$, Fig.~\ref{fig: sample_3}(e)). 
However, when the virtual particle deposition engine is active, the dynamic co-deposition of calculated corrector spheres acts as a temporary mechanical shield, preventing over-compaction and preserving the uncompacted, open pore pathways. 
Following the computational removal of these spheres, the final reconstructed model (Fig.~\ref{fig: sample_3}(f)) achieved a target porosity of $\phi = 0.3325$, matching the true experimental target ($\phi = 0.3319$) with measured absolute deviation just $0.0006$ ($0.06\%$).

The morphological validation is verified by the study of the two-point correlation function $S_2(r)$, shown in Fig.~\ref{fig: sample_3_dist}(c). 
In contrast to the previous synthetic cases where 3D structural descriptors were utilized, the validation here is conducted strictly using 2D data pools to ensure mathematical consistency with the baseline SEM slice. 
The statistical clouds (ensembles) for both the benchmark XCT and our reconstructed framework were generated by computing 20 individual $S_2(r)$ curves across a multitude of distinct 2D random cross-sectional slices extracted from their respective 3D models. 
These ensembles were plotted against a single baseline $S_2(r)$ curve computed directly from the uncorrupted 2D SEM image.

The results demonstrate that the baseline curve derived from the real SEM data falls strictly within the statistical cloud of our physically reconstructed 2D slices. Conversely, the $S_2(r)$ ensemble from the XCT volume experiences a notable spatial deviation, pulled away by the phantom internal porosity.

This geometric alignment translates directly into hydrodynamic accuracy: the absolute permeability calculated from the pore-network model of the benchmark XCT volume is $K = 1284~\text{mD}$, while our final reconstructed model yields a stable, artifact-free absolute permeability of $K = 1183~\text{mD}$. 
This minor deviation ($\sim 8\%$) proves that our framework successfully suppresses systematic tomographic artifacts, providing a robust alternative for digital rock twins where conventional XCT workflows fail.

\section{General Discussion}

The shift from purely geometric optimization to a physically consistent synthesis framework marks a major paradigm change in the 3D digital reconstruction of multi-scale, heterogeneous media from 2D images. 
A direct comparative analysis with our previous work on this topic~\cite{kulygin2024pore} reveals the critical limitations of that earlier methodology and highlights the transformative capabilities of the current framework. 

In our previous algorithm, the reconstruction of the grain backbone was fundamentally non-physics-based. 
Because it relied on purely geometric shuffling and sequential particle expansion to eliminate structural overlaps, it imposed severe, rigid constraints on the simulation output. 
First, that earlier framework was inherently restricted to simplified grain geometries and could only function within a narrow, low-porosity range. 
Second, it possessed no capacity to control multi-scale void spaces or preserve large, non-equilibrium structural features. 
Third, it was strictly limited to a single amorphous phase, whereas real heterogeneous materials routinely feature multiple distinct cementing components with complex mutual layout rules. 

The physically driven framework introduced in this Part 1 study completely breaks through these technological barriers by utilizing a high-performance rigid-body dynamics engine (NVIDIA PhysX). 
Instead of relying on artificial geometric manipulations, every generated grain together with special virtual particles settle into local minima of the gravitational potential based on Newtonian contact laws. 
This fundamentally eliminates the old constraints on porosity ranges, enabling the high-fidelity synthesis of digital twins across a vastly broader spectrum of synthetic and natural objects---ranging from highly porous, weathered matrices to complex multi-phase systems.

This complete replacement of blurred tomographic voxels and non-physical geometric shuffling with an authentic physical grain backbone unlocks critical sub-micron and macroscale insights that traditional digital rock physics pipelines fail to capture:

\begin{enumerate}
    \item First, the framework provides physically plausible grain-to-grain contact networks. 
    Having access to explicit contact areas and realistic coordination numbers is an absolute prerequisite for advanced multiscale modeling. 
    It provides the necessary structural foundation to accurately simulate rock and material deformation under stress, inelastic structural alterations, and effective thermal conductivity tensors. 
    This is particularly vital for evaluating weakly consolidated porous media, where laboratory testing is heavily compromised. 
    In routine physical laboratory core-flooding or triaxial tests, these fragile formations experience intense mechanical deformations. 
    Applying even minor load pressures (under 20~MPa) to simulate reservoir conditions induces grain displacement, which triggers a systematic 5--15\% deviation in empirical permeability measurements. 
    Our workflow circumvents this experimental pitfall by numerically generating an uncorrupted, uncompacted initial state of the matrix.
    \item Second, the method ensures precise, deterministic control over the spatial distribution of different mineral and phase fractions within the 3D volume. 
    In the context of multi-phase transport, knowing exactly where each phase resides in 3D space allows researchers to populate the digital twin with highly realistic, heterogeneous wettability maps. 
    These local chemical variations serve as the primary driving factor governing two-phase fluid displacement, capillary pressure curves, and relative permeability behavior---critical parameters that conventional XCT workflows cannot evaluate due to the lack of phase contrast between chemically distinct components.
    \item Third, our introduction of a two-stage deposition pipeline utilizing temporary virtual corrector spheres resolves the physical compaction problem. 
    A pure Newtonian settling of non-convex grains inevitably minimizes the pore space volume, automatically driving the granular skeleton toward its tightest compaction state. 
    By acting as transient mechanical shields during active physical settling, these corrector spheres dynamically protect large, non-equilibrium void architectures (such as dissolution channels or structural macroscopic pores) from being choked by smaller particles. 
    Once mechanical equilibrium is achieved, the spheres are computationally removed, yielding a physically stable matrix that honors target 2D morphological descriptors with an exceptional absolute porosity tolerance of 0.02\%--0.06\%.
\end{enumerate}

Crucially, the successful deployment of this physics-driven framework establishes a highly transferable 2D to 3D reconstruction paradigm that extends well beyond the domain of digital rock physics into broader materials science. By replacing abstract geometric shuffling with authentic Newtonian mechanics, the developed pipeline provides a rigorous foundation for predicting the macroscopic behavior of any disordered particulate or granular matter governed by multiscale pore-throat networks. For instance, in granular engineering and powder metallurgy, the explicitly resolved grain-to-grain contact networks and coordination numbers serve as vital structural inputs for mechanics models where macroscopic compaction and tensile properties are sensitive to localized force networks. Similarly, in advanced energy storage applications, such as the synthesis of lithium-ion battery electrodes from cross-sectional micrographs, the ability to accurately decouple multi-phase boundaries—including active material particles, binders, and interconnected porous domains—is paramount for capturing anomalous ionic transport and minimizing tortuosity anomalies \cite{yang2026high, grafensteiner2025data, jang2025digital}. Furthermore, the dual-stage packing logic developed herein holds direct relevance for additive manufacturing feedstocks and architectural composites, where optimized particle-size blends and precise porosity regulation dictate the rheology, densification, and ultimate mechanical performance of additively manufactured ceramics and printed concrete \cite{meier2025surface, liu2025dlp}. By formulating the 2D to 3D inversion problem around stationary morphological invariants and real-time rigid-body kinetics, this framework operationalizes a universal bridge from sparse two-dimensional microstructural data to hydrodynamically and mechanically equivalent three-dimensional digital twins across diverse porous media.

Furthermore, this physical stability opens direct industrial pathways for the development of automated petrophysical and materials characterization software. 
Because the framework can synthesize hydrodynamically equivalent 3D twins directly from uncorrupted 2D images, it is uniquely suited for evaluating highly damaged or completely destroyed granular matrices. It enables the accurate prediction of material properties from fine rock fragments, drill cuttings, industrial powders, or heavily crushed unconsolidated samples where extracting a standard cylindrical core plug or manufacturing a macroscopic sample for micro-XCT scanning is physically impossible. 

As this paper represents Part 1 of an interconnected research cycle, the established gravitational deposition framework serves as the structural foundation for upcoming technological expansions. 
While this work successfully controls porosity and size distributions via simplified spherical sub-particles, Part 2 of this series will expand the geometric frontier by introducing advanced algorithms for non-convex 3D grain shape generation and stochastic ensembles that precisely honor multi-dimensional morphological shape descriptors derived from 2D masks. 
Subsequently, Part 3 will advance the statistical theory of random media under strict spatial confinement, introducing the mathematical machinery to replace the virtual corrector spheres with multiple, distinct, interconnected amorphous cementing phases and clay gels, honoring their mutual structural constraints and spatial positioning.

\section{Conclusion } \label{Conclusion}

This work presents a physically consistent framework for reconstructing three-dimensional granular media from high-resolution two-dimensional image data. By integrating stereological inversion, grain-shape reconstruction, and a two-stage gravitational packing strategy with temporary virtual spheres, the method establishes a direct link between observable 2D structural features and 3D reconstructed architectures. The resulting digital reconstructions preserve the target grain morphology, recover the corrected size distribution, and reproduce the essential pore-space statistics that govern transport behavior.

The proposed pipeline is validated on synthetic and natural benchmark systems, demonstrating that the reconstructed samples reproduce porosity, pore-size distributions, and two-point correlation functions with high fidelity. In particular, the framework remains robust when conventional XCT workflows fail because of partial-volume effects, edge compaction, or sub-resolution porosity. The comparative analysis shows that neglecting stereological correction leads to systematic distortions in the grain backbone and to substantial underestimation of permeability, even when the total porosity is matched. This finding highlights the importance of physically grounded reconstruction over purely geometric or volumetric matching.

More broadly, the framework provides a practical methodology for generating digital twins of heterogeneous porous media in cases where direct 3D acquisition is limited, corrupted, or unavailable. The approach is especially relevant for weakly consolidated sandstones, dissolution-affected materials, and other granular systems whose microstructure is strongly controlled by non-equilibrium packing and contact mechanics. The results therefore establish a foundation for future extensions toward non-convex grain generation, multi-phase cementing structures, and more complete stochastic representations of random media.

Overall, this study demonstrates that a physics-informed reconstruction strategy can recover statistically accurate and hydrodynamically meaningful 3D pore structures from 2D data, offering a viable route for digital rock modeling and related materials-science applications.

\section*{ACKNOWLEDGEMENTS}
This work was supported by the Russian Science Foundation (Grant No. 25-13-00313).

\nocite{*}
\bibliography{references.bib}

\section{Appendix} \label{Appendix}

As described in the Methods section, the gravitational sedimentation simulations were performed with the PhysX engine for all cases using the same parameter set. This ensured that differences between the reconstructed samples were attributable to the input microstructures rather than to variations in the numerical setup. The principal parameters governing the sedimentation process are summarized in Table \ref{table: table}. The simulation was terminated when all particles satisfied the PhysX sleep criterion, i.e., when their translational velocity remained below the prescribed threshold for more than 300 consecutive simulation steps.

\onecolumngrid

\begin{table}[!htp]
\center
\raggedright
  \centering
  \footnotesize\centering
    \captionsetup{justification=centering}
    \captionsetup{size=footnotesize}
    \caption{Main physical simulation parameters used in the PhysX-based implementation. }

    \bgroup
    \def\arraystretch{1.9}
    \setlength\tabcolsep{9.5 pt}
    \begin{tabular}{|c|c|c|c|}
    \hline
    Parameter & Class & Method & Value \\
    \hline
    Gravity & \texttt{PxSceneDesc} & \texttt{gravity} & $9.8$ \\
    \hline
    Simulation time step & \texttt{PxScene} & \texttt{simulate} & $0.01$ \\
    \hline
    Static friction & \texttt{PxMaterial} & \texttt{staticFriction} & $0.5$ \\
    \hline
    Dynamic friction & \texttt{PxMaterial} & \texttt{dynamicFriction} & $0.5$ \\
    \hline
    Elasticity / restitution & \texttt{PxMaterial} & \texttt{restitution} & $0.5$ \\
    \hline
    Mass & \texttt{PxRigidDynamic} & \texttt{setMass} & \parbox[c]{5cm}{Recomputed for each grain from the material density $2.6$} \rule[-3.0ex]{0pt}{7.0ex}\\
    \hline
    Inertia & \texttt{PxRigidDynamic} & \texttt{setMassSpaceInertiaTensor} & Diagonal tensor $(1.0, 1.0, 1.0)$ \\
    \hline

    Fixation in a stationary state & \texttt{PxRigidDynamic} & \texttt{putToSleep} & \parbox[c]{5cm}{If a body remains below a velocity  threshold of $0.01$ for more than $300$ simulation steps, it is fixed in place} \rule[-5.0ex]{0pt}{11.5ex}\\
    \hline
    \end{tabular}
    \egroup
    
  \label{table: table}
\end{table}

\twocolumngrid

\end{document}